\documentclass[11pt, letterpaper]{article}

\usepackage{graphicx}
\usepackage{amsmath}
\usepackage{amssymb}
\usepackage{setspace}
\usepackage{epstopdf}
\usepackage{color}
\usepackage[letterpaper,left=1in,right=1in,top=1in,bottom=1in]{geometry}
\usepackage[linesnumbered,ruled,vlined,longend]{algorithm2e}
\usepackage{multirow}
\usepackage{longtable}
\usepackage{rotating}
\usepackage{threeparttable}
\usepackage{colortbl}
\usepackage{amsthm}
\usepackage[modulo]{lineno}
\usepackage{enumerate}
\usepackage{bm}
\usepackage{algorithmic}

\newtheorem{remark}{Remark}

\theoremstyle{definition}

\title{\Large \bf Economic model predictive control of integrated energy systems: A multi-time-scale framework}

\author{\centerline{\normalsize Long Wu$^{a,b}$, Xunyuan Yin$^{c}$, Lei Pan$^{a,\ast}$, Jinfeng Liu$^{b,}$\thanks{Corresponding author: L. Pan. Email: panlei@seu.edu.cn; J. Liu. Tel: +1-780-492-1317. Fax: +1-780-492-2881. Email: jinfeng@ualberta.ca}}\vspace{5mm}\\
    \centerline{\small $^{a}$ National Engineering Research Center of Power Generation Control and Safety,}\\
    \centerline{\small  School of Energy and Environment, Southeast University, Nanjing, 210096, China}\\
    \centerline{\small $^{b}$ Department of Chemical \& Materials Engineering, University of Alberta,}\\
    \centerline{\small Edmonton, Alberta, Canada, T6G 1H9}\\
    \centerline{\small $^{c}$ School of Chemical and Biomedical Engineering, Nanyang Technological University,}\\
    \centerline{\small 62 Nanyang Drive, Singapore, 637459}}
    
\allowdisplaybreaks

\begin{document}

\date{}

\maketitle
\setstretch{1.39}

\begin{abstract}
In this work, a composite economic model predictive control (CEMPC) is proposed for the optimal operation of a stand-alone integrated energy system (IES), aiming at meeting customers' electric and cooling requirements under diverse environmental conditions while reducing fuel consumption. Time-scale multiplicity exists in IESs dynamics is taken into account and addressed using multi-time-scale decomposition based on singular perturbation theory. Through multi-time-scale decomposition, the entire IES is decomposed into three reduced-order subsystems with slow, medium, and fast dynamics. Subsequently, the CEMPC, which includes slow economic model predictive control (EMPC), medium EMPC and fast EMPC, is developed to match the time-scale multiplicity featured in IESs dynamics. The EMPCs communicate with each other to ensure consistency in decision-making. In the slow EMPC, the global control objectives are optimized, and the manipulated inputs explicitly affecting the slow dynamics are applied. The medium EMPC optimizes the control objectives correlated with the medium dynamics and applies the corresponding optimal medium inputs to the IES, while the fast EMPC optimizes the fast dynamics relevant objectives and makes a decision on the manipulated inputs directly associated with the fast dynamics. Meanwhile, thermal comfort is integrated into the CEMPC in the form of zone tracking of the building temperature for achieving more control degrees of freedom to prioritize satisfying the electric demand and reducing operating costs of the IES. Moreover, a long-term EMPC based on a simplified slow subsystem model is developed and incorporated into the CEMPC to ensure that the operating state accommodates long-term forecasts for external conditions. Finally, the effectiveness and superiority of the proposed method are demonstrated via simulations and a comparison with a hierarchical real-time optimization mechanism.
\end{abstract}


\section{Introduction}
With the rise of concern about climate change and energy shortage, clean energy development has been highly promoted. Renewable energy and natural gas are playing an increasingly important role in electric power supply networks worldwide \cite{r1}. In this case, there have been significant demands for integrated energy systems (IESs) because of their potential of absorbing renewable energy and improving overall fuel efficiency in distributed energy systems. However, IESs have many characteristics different from traditional coal-fired power plants. Traditional centralized coal-fired power plants usually provide electricity to customers through electrical grids in a single direction. In contrast, IESs and customers are more directly interconnected via microgrids and pipelines. With penetration of intermittent renewable energy, such as solar and wind energy, IESs are more susceptible to external environments. Thus, the operating state of IESs has to be optimized frequently to satisfy customers' demands and keep reliable and economical operations under varying external conditions. The most crucial feature of IESs is that IESs are usually composed of various operating unis that are coupled through the working substance and energy flows to provide electricity as well as cooling or heating, or both, to customers, which makes IESs display diverse dynamics and possess the time-scale multiplicity. These factors are challenging for conventional control schemes.

Most of the existing results on control of the IESs focus on the long-term operation optimization. In particular, a considerable amount of literature has dealt with the long-term operating strategy from various aspects. For example, the peak load shifting potential of energy storage and heating network was adopted to improve system flexibility and economics in \cite{r3}, \cite{r4}. In \cite{r5,r6,r7}, the environmental and demand uncertainty was taken into account to optimize the long-term operating state and daily operating cost of IESs. Two-layer optimization schemes were extensively investigated in \cite{r8,r9,r10,r11,r51} from the perspective of improving the profitability and maintaining the energy balance in distributed energy systems. In \cite{r53}, the optimal scheduling problem of a CCHP microgrid system was artificially separated into three stages, and then a hierarchical optimization framework was adopted to handle these sub-problems. In \cite{r12}, the floating price of electricity and the interaction between system and power grid were employed to increase profitability and reduce negative effects on the power grid. Most of the mentioned works address the optimization problem by constructing a steady-state mixed-integer linear program, while a few pieces of work take into consideration the enormous thermal inertia of building, see \cite{r14,r15,r16}. Moreover, a two-level model predictive control scheme that takes into account part of the dynamic behavior of equipment was developed in \cite{r13} for accelerating system response. Model predictive control (MPC) was adopted in \cite{r3,r51,r13,r53,r15,r16,r54} to enhance systems' robustness due to its receding horizon implement and feedback correction. However, it is noted that humans' insensitivity to variations in building temperature within a small range, the so-called thermal comfort \cite{r17}, was not taken into account explicitly to improve electric load tracking performance in most IESs relevant studies, except in \cite{r11,r15,r16}. Furthermore, the existing studies have in general considered the long-term operation optimization and the short-term optimization independently. This causes the system to operate at a suboptimal mode.

On the other hand, for IESs or distributed renewable energy systems, a small amount of literature is devoted to designing short-term control strategies or coordinated control schemes that consider complete dynamics in operating units. The limited attention was mainly given to hybrid energy systems with relatively simple system structures. Specifically, supervisory and distributed predictive control methods were first proposed in \cite{r18}, \cite{r19} for wind-solar hybrid energy systems, and similar hierarchical approaches were also found in \cite{r20}, \cite{r21} later. Supervisory predictive control and adaptive control were also developed for hybrid energy systems consisting of microturbine, solar, wind turbine, and batteries, see \cite{r22}, \cite{r23}. A two-time-scale linear quadratic controller was proposed to balance real-time electric power in a small-scale linear smart grid \cite{r56}. In \cite{r52}, a distributed MPC was developed for grid-interactive buildings to alleviate communication burden and balance competition between electric power and building temperature tracking. In addition, sliding mode control has been discussed in a hybrid energy system composed of solar and fuel cells \cite{r24}. However, little attention has been given to a design of short-term or coordinated control strategies for large-scale complex nonlinear IESs. Additionally, the applications of centralized economic MPC (EMPC) in energy systems have been reported in \cite{r26}, \cite{r27}, which developed a new control strategy for coal-fired power plant and post-combustion carbon capture process, respectively. We note that EMPC of great ability to simultaneously handle optimization and tracking problems is in advance of conventional hierarchical control schemes, which is highly appropriate for IESs coordinated control problem. However, the attention was scarcely paid to that. To sum up, although the existing efforts contributed to achieving the coordinated control of distributed renewable energy systems, most of these studied simple hybrid energy systems do not have as many energy forms, time-scale multiplicity and strong couplings as IESs, such that the proposed control schemes are no longer suitable for large-scale complex IESs.

From the existing works, it can be seen that there are plenty of achievements of the IES long-term optimization. However, less effort was made to design short-term coordinated control schemes for IESs, but that is of critical importance to the optimal operation of IESs. For the design of short-term control schemes for IESs, there are two intuitive ideas. The first is to design a controller for each operating unit in a hierarchical framework. But then, the strong couplings in IESs and the possibility of using them to improve systems' performance in transient processes are almost ignored such that control performance degradation is inevitable. This solution will also increase the cost of a communication network. The second is to design a centralized controller in the hierarchical framework for the entire system since it can precisely capture systems' couplings and provide better performance comparing the first strategy \cite{r55}. However, the issue of this scheme is that the centralized structure will lead to a unmanageable computation complexity and cannot deal with the time-scale multiplicity in IESs' dynamics, which will cause an ill-conditioned control problem or the loss of closed-loop stability.

These considerations motivate us to propose a novel composite economic MPC based on the multi-time-scale decomposition, inspired by \cite{r28,r29,r30}, for large-scale IESs in this work. First of all, the time-scale multiplicity exhibited in the dynamics of a stand-alone IES is investigated. Then, by using the multi-time-scale separation based on singular perturbation theory, the stand-alone IES is decomposed into three reduced-order subsystems: slow, medium, and fast subsystem. Next, according to each subsystem's dynamics, a composite economic MPC (CEMPC) including three interconnected EMPCs, a slow EMPC, a medium EMPC, and a fast EMPC, is developed, in which communication will ensure consistency in optimization results. In terms of the slow EMPC, the controller is designed based on the slow dynamics, and the global control objectives, satisfying the customers' demands on the electricity and cooling while improving the system's profitability, are optimized. The manipulated inputs explicitly correlated with the slow dynamics are applied to the IES here. Meanwhile, a zone tracking framework for building temperature is integrated into the slow EMPC via a zone tracking cost to realize the mentioned thermal comfort and improve the electric power tracking performance and system profitability. Subsequently, the control objectives that is relevant to the medium dynamics, satisfying the electric demands and improving the system's profitability, and optimal references from the slow EMPC are considered in the medium EMPC based on the medium subsystem to optimize the corresponding medium inputs and implement them. Further, the fast subsystem is used to formulate the fast EMPC, which optimizes the inputs and control objectives directly associated with the fast dynamics while considering the optimal reference from the medium EMPC. Furthermore, to ensure the started operating units can satisfy the customers' demands and energy storage units adapt to the periodic customers' demands and environmental conditions, a long-term EMPC with zone tracking based on a simplified slow subsystem model is designed and incorporated in the CEMPC to optimize start-ups/shut-downs of the operating units and the operational curves of energy storage units. The thermal comfort is also taken into account in the long-term EMPC. Finally, a hierarchical real-time optimization is also presented for comparison purposes.

Although there are results of the short-term coordinated control for hybrid energy systems, this work is the first to develop a coordinated control scheme for large-scale complex IESs by considering the multi-time-scale subsystem decomposition. The contributions of this work are as follows:

(1) The dynamic time-scale multiplicity in IESs is investigated and addressed with the multi-time-scale subsystem decomposition based on singular perturbation theory, by which the entire system is decomposed into three reduced-order subsystems, and control objectives are decomposed correspondingly.

(2) A composite economic MPC based on the subsystem decomposition is proposed for achieving optimal coordinated operation of IESs, capturing the multi-time-scale dynamics, lightening the computational burden, dealing with the couplings, and simultaneously handling optimization and tracking problems.

(3) The thermal comfort is integrated into the proposed CEMPC by exploiting the zone tracking of building temperature to give IESs freedom from the competition between electrical and cooling tracking objectives to prioritize satisfying the electric demand and improve the economics of IESs.

The remainder of this paper is organized as follows: in Section 2, a stand-alone IES for cooling and electricity supply and its dynamic model are introduced, then the control problem that will be tackled is formulated; Section 3 investigates the time-scale multiplicity in the IES and presents the system decomposition based on time-scale multiplicity; Section 4 presents the CEMPC design comprises the long-term EMPC and the short-term EMPC; simulations are performed to demonstrate the effectiveness and superiority of the proposed method in Section 5; conclusions are drawn in Section 6.

\section{System description and control problem formulation}
\subsection{System description}
\begin{figure}[t]
	\centering
	\includegraphics[width=0.92\hsize]{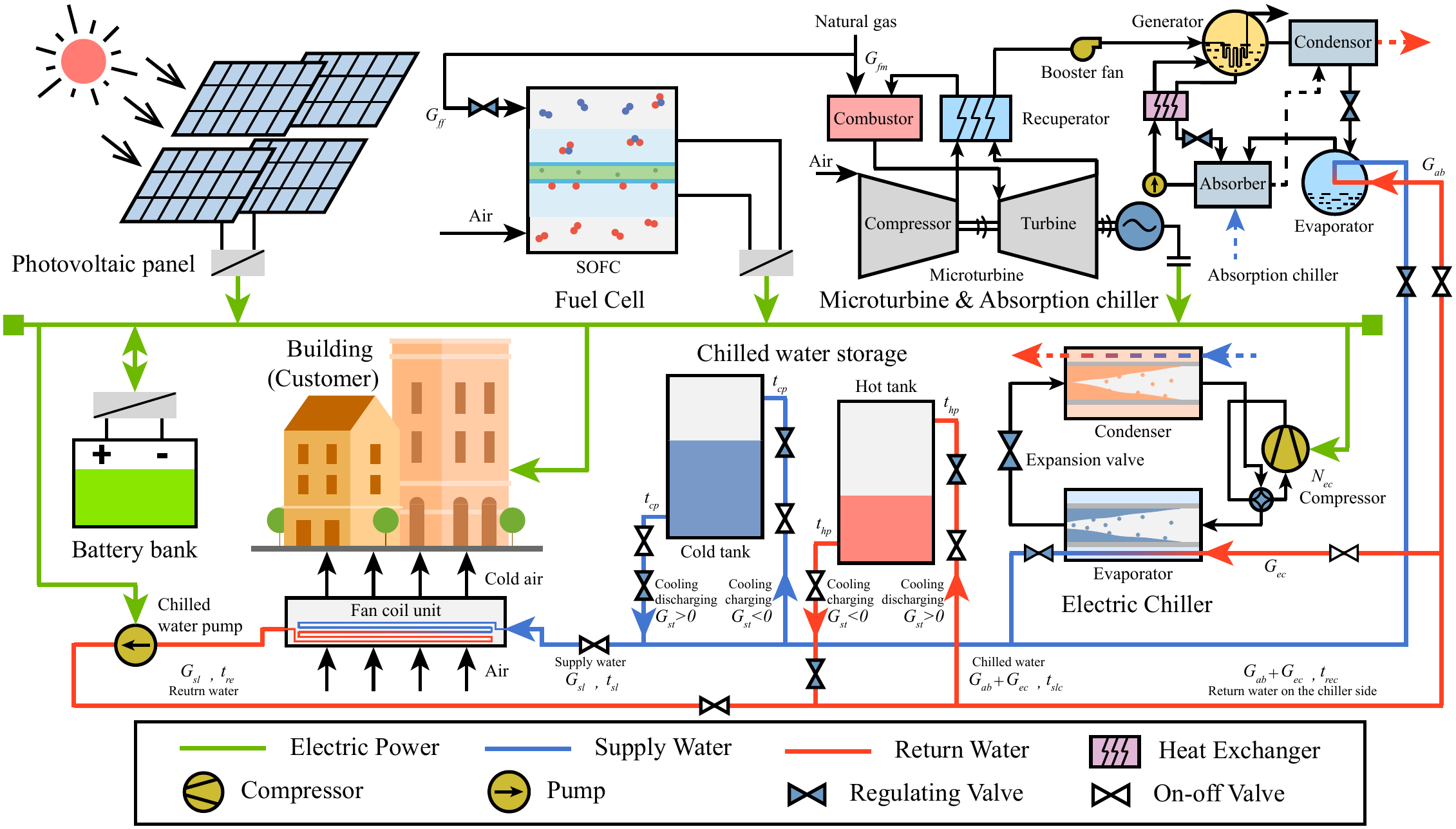}
	\caption{The considered integrated energy system}
	\label{f1}
\end{figure}

A schematic diagram of the IES studied in this work is shown in Figure \ref{f1}. The system is operated in the stand-alone mode, providing customers with electricity and cooling via microgrid and pipeline, where the cooling is utilized to regulate the building temperature. The IES consists of seven operating units: a photovoltaic energy generation unit, a fuel cell, a microturbine integrated with an absorption chiller, an electric chiller, a chilled water storage unit, a building, and a battery bank. The photovoltaic energy unit generates electricity directly from sunlight via an electronic process, but it is susceptible to environmental conditions. Therefore, the fuel cell and microturbine are introduced into the IES. In the microturbine, the natural gas is directly fed into the combustor, which produces high temperature and pressure flue gas to drive the turbine and generator to convert chemical energy into electric energy, while the fuel cell utilizes hydrogen reformed from natural gas to react with oxygen for generating electric power. Meanwhile, the battery bank is adopted to keep the real-time electric power balance in the stand-alone microgrid and keep pace with periodic changes in renewable energy and customers' demands to absorb renewable energy and improve system efficiency. On the other hand, the waste heat from the combustor in the microturbine is sent to the downstream absorption chiller for heat transfer to the lithium bromide solution to generate chilled water, while the electric chiller produces chilled water by using the electricity to circulate refrigerant in the evaporator and condenser. Then the chilled water from the absorption chiller and electric chiller is mixed and piped to the customers and/or cold tank via the water pump. In the meantime, the chilled water storage unit is employed to achieve peak cooling load shifting and improve the flexibility and economics of the IES. In the cooling discharging mode, the cold water is discharged from the cold tank and mixed with the chilled water from the chillers to form the supply water that is piped into the fan coil units for cooling air in the building. After heat transfer, the return water is sent to the chillers and the hot tank, respectively. In the cooling charging mode, a portion of the chilled water from the chillers is piped to the customers, and the remainder of the chilled water is sent into the cold tank while the hot water is discharged from the hot tank and mixed with the return water from the customers to go to the chillers for cooling.

\subsubsection{Photovoltaic energy generation unit}
The photovoltaic energy generation unit uses solar energy to generate electric power, which can help achieve a higher overall fuel efficiency of the IES. We note that the model of photovoltaic panel is usually in steady-state form and we assume that the maximum peak power tracking is consistently implemented to maximally utilize the solar energy \cite{r33}. The current and the voltage of each photovoltaic cell at the maximum power point, $I_{pv,max}$ and $V_{pv,max}$, can be modeled as algebraic functions of the ambient temperature and the solar radiation as follows \cite{r39}, \cite{r40}:
\begin{equation}
	I_{pv,max} = I_{max,0}\frac{S_{ra}}{S_{ra,0}}(1 + c_{pv,1}(t_a - t_{pv,0})) \label{eq_add_1}
\end{equation}
\begin{equation}
	V_{pv,max} = V_{max,0}(1 - c_{pv,3}(t_a - t_{pv,0}))\ln (e + c_{pv,2}(S_{ra} - S_{ra,0})) \label{1}
\end{equation}
where $I_{max,0}$ and $V_{max,0}$ are the current and voltage of each photovoltaic cell at the maximum power point under nominal environmental condition; $S_{ra,0}$ and $t_{pv,0}$ are the nominal solar radiation and ambient temperature; $S_{ra}$ and $t_{a}$ represent the actual solar radiation and ambient temperature; $c_{pv,i}, \ i=1,2,3$ are the coefficients; $e$ is Euler's number. Then the electric power generated by the photovoltaic panel $P_{pv}$ can be obtained from: 
\begin{equation}
	P_{pv} = n_{pp}n_{sp}\frac{I_{pv,max}V_{pv,max}}{1000}
\end{equation}
where $n_{pp}$ and $n_{sp}$ are the number of photovoltaic cells in parallel and series, respectively.

\subsubsection{Fuel cell}
Fuel cells are viewed as promising technologies for future clean energy. Since fuel cells have a flexible operation mode, they are used in the IES to complement photovoltaic energy unit and microturbine. A benchmark model of solid oxide fuel cells described in \cite{r34}, \cite{r35} is employed in the IES. The dynamics of the fuel cell are as follows:
\begin{equation}
	\frac{d I_f}{d\tau} = z_{fc}\frac{I_{fc} - I_f}{\tau_e} \label{4}
\end{equation}
\begin{equation}
	\frac{d G_{H_2}}{d\tau} = z_{fc}\frac{M_{ng}G_{ff} - G_{H_2}}{\tau_f}
\end{equation}
\begin{equation}
	\frac{d p_{O_2}}{d\tau} = z_{fc}\frac{\frac{G_{O_2} - I_f K_r}{K_{O_2}} - p_{O_2}}{\tau_{O_2}}
\end{equation}
\begin{equation}
	\frac{d p_{H_2O}}{d\tau} = z_{fc}\frac{\frac{G_{H_2r}}{K_{H_2O}} - p_{H_2O}}{\tau_{H_2O}}
\end{equation}
\begin{equation}
	\frac{d p_{H_2}}{d\tau} = z_{fc}\frac{\frac{G_{H_2} - G_{H_2r}}{K_{H_2}} - p_{H_2}}{\tau_{H_2}} \label{8}
\end{equation}
where $I_{f}$ and $I_{fc}$ are the inside and outside currents of the fuel cell; $z_{fc}$ is an integer variable representing switching fuel cell on/off, specifically, the fuel cell is switched on when $z_{fc}=1$, otherwise, the fuel cell is switched off, and $z_{fc}=0$; $\tau_e$ is the corresponding time constant; it is assumed that $I_{fc}$ can be regulated by the converter for keeping fuel utilization \cite{r46}, then $I_{fc}$ satisfies $I_{fc} = G_{H_2}fu_d/(2K_r)$ where $K_r$ is a coefficient and $fu_d$ is the desired fuel utilization as a constant; $G_{H_2}$ represents the molar flow rate of hydrogen reformed from the natural gas $G_{ff}$ that is a manipulated input; $M_{ng}$ is the conversion factor for the molar mass of natural gas; $\tau_e$ is the time constant; $p_{O_2}$ denotes the partial pressure of oxygen; $G_{O_2}$ is the oxygen flow rate depending on $G_{O_2} = G_{H_2}/r_{H-O}$ where $r_{H-O}$ is the ratio of hydrogen to oxygen; $K_{O_2}$ is a constant factor; $\tau_{O_2}$ is the time constant of state $p_{O_2}$; $p_{H_2O}$ denotes the partial pressure of vapor; $K_{H_2O}$ and $\tau_{H_2O}$ are the factor and time constant, respectively; $p_{H_2}$ represents the partial pressure of hydrogen; $G_{H_2r}$ is the reacted hydrogen evaluated by $	G_{H_2r} = 2K_r I_f$; $K_{H_2}$ and $\tau_{H_2}$ are the corresponding factor and time constant, respectively. The potential voltage of fuel cell is expressed below:
\begin{equation}
	V_0 = N_{0fc} \left(E_{0fc} + \frac{R_{0fc}T_{0fc}}{2F_0}\ln \frac{p_{H2}\sqrt{p_{O2}}}{p_{H2O}} \right)
\end{equation}
The output voltage of fuel cell $V_{fc}$ is presented in $V_{fc} = V_0 - \eta_a - \eta_{c} - \eta_{o}$ where $\eta_a, \ \eta_{c}, \ \eta_{o}$ are the voltage losses for activation, concentration and resistance, evaluated by $\eta_a=a_{fc}+b_{fc}\ln I_f$, $\eta_{c}=-R_{0fc}T_{0fc}\ln (1-I_f/I_L)/(2F_0)$ and $\eta_{o}=I_f r_{fc}$. The electric power of fuel cell, $P_{fc}$, is computed as follows:
\begin{equation}
	P_{fc} = z_{fc}\frac{V_{fc}I_{fc}}{10^3}
\end{equation}

\subsubsection{Microturbine integrated with absorption chiller}
The microturbine integrated with an absorption chiller plays a critical role in the IES due to its high reliability and efficiency. The dynamics of a microturbine integrated with an absorption chiller generation system can be described by a simplified model as follows \cite{r36}:
\begin{equation}
	\frac{d P_{mtf}}{d\tau} = z_{ma} \frac{k_{ma,1}G_{fm}^2+k_{ma,2}G_{fm}+k_{ma,3}-P_{mtf}}{\tau_{mtf}} \label{11}
\end{equation}
\begin{equation}
	\frac{d t_{abf}}{d\tau} = z_{ma} \frac{k_{ma,4}P_{mtf}-t_{abf}}{\tau_{abf}}
\end{equation}
\begin{equation}
	\frac{d t_{abw}}{d\tau} = z_{ma} \frac{k_{ma,5}G_{ab}+k_{ma,6}-t_{abw}}{\tau_{abw}}
\end{equation}
\begin{equation}
	\frac{d t_{abt}}{d\tau} = z_{ma} \frac{k_{ma,7}t_{rec}+k_{ma,8}-t_{abt}}{\tau_{abt}} \label{14}
\end{equation}
where $P_{mtf}$ characterizes the effect of the natural gas flow rate $G_{fm}$ on the electric power produced by the microturbine; $z_{ma}$ is a 0-1 integer variable: the microturbine integrated with absorption chiller is turned on if $z_{ma}=1$, otherwise, it is turned off; $k_{ma,i}, \ i=1,\dots,8$ in the dynamics are constant factors; $\tau_{mtf}$ is the time constants of the state $P_{mtf}$; $t_{abf}$ describes the microturbine's influence on the supply chilled water temperature in the downstream absorption chiller; $\tau_{abf}$ is the corresponding time constant; $t_{abw}$ reflects the effect of the chilled water flow rate $G_{ab}$ on the supply chilled water temperature; $\tau_{abw}$ is the time constant of the state $t_{abw}$; $t_{abt}$ represents the effect on the supply chilled water temperature caused by the return water temperature on the chiller side $t_{rec}$; the detailed expression of $t_{rec}$ will be given in Section 2.1.6; $\tau_{abt}$ is the time constant of $t_{abt}$. The electric power generated by the microturbine $P_{mt}$ and the supply chilled water temperature produced by the absorption chiller $t_{ab}$ can be obtained from:
\begin{equation}
	P_{mt} = z_{ma} (P_{mt,0} + P_{mtf})
\end{equation}
\begin{equation}
	t_{ab} = z_{ma} (t_{ab,0} + t_{abf} + t_{abw} + t_{abt})
\end{equation}
where $P_{mtf,0}$ and $t_{ab,0}$ are the nominal power and supply chilled water temperature. Then, the cooling power provided by the absorption chiller can be achieved from $Q_{ab} = G_{ab}C_{w}(t_{rec} - t_{ab})$, where $C_{w}$ is the specific heat capacity of water. Note that we assume all of the waste heat from the microturbine is sent to the absorption chiller. In the microturbine with an absorption chiller generation system, $G_{fm}$ and $G_{ab}$ are manipulated variables to regulate $P_{mt}$ and $t_{ab}$. For a more detailed explanation and complete description of the microturbine integrated with absorption chiller, please refer to \cite{r36}.

\subsubsection{Electric chiller}
The electric chillers uses electricity to provide cooling for customers as a supplement to the absorption chiller in the IES. The dynamic model of the electric chiller, developed in \cite{r41}, \cite{r42}, is as follows:
\begin{equation}
	\frac{d t_e}{d\tau} = z_{ec}\frac{G_r(h_{eri} - h_{ero}) + \alpha_r A_{ei}(t_{es} - t_e)}{C_{er}M_{er}} \label{17}
\end{equation}
\begin{equation}
	\frac{d t_{es}}{d\tau} = z_{ec}\frac{\alpha_r A_{ei}(t_e - t_{es}) + \alpha_w A_{eo}(t_{ewm} - t_{es})}{C_s M_{es}}
\end{equation}
\begin{equation}
	\frac{d t_{ewm}}{d\tau} = z_{ec}\frac{C_w G_{ec}(t_{rec} - t_{ec}) + \alpha_w A_{eo}(t_{es} - t_{ewm})}{C_w M_{ew}}
\end{equation}
\begin{equation}
	\frac{d t_c}{d\tau} = z_{ec}\frac{G_r(h_{cri} - h_{cro}) + \alpha_r A_{ci}(t_{cs} - t_c)}{C_{cr}M_{cr}}
\end{equation}
\begin{equation}
	\frac{d t_{cs}}{d\tau} = z_{ec}\frac{\alpha_r A_{ci}(t_c - t_{cs}) + \alpha_w A_{co}(t_{cwm} - t_{cs})}{C_s M_{cs}}
\end{equation}
\begin{equation}
	\frac{d t_{cwm}}{d\tau} = z_{ec}\frac{C_w G_{cw}(t_{cwi} - t_{cwo}) + \alpha_w A_{co}(t_{cs} - t_{cwm})}{C_w M_{cw}} \label{22}
\end{equation}
where $t_e$ denotes the evaporating temperature in the evaporator; $z_{ec}$ indicates switching electric chiller on/off by setting $z_{ec}=1$ and $z_{ec}=0$; $G_r$ is the the mass flow rate of refrigerant and will given later; $h_{eri}$ and $h_{ero}$ represent the enthalpy of refrigerant at the evaporator inlet and outlet, respectively; $\alpha_r$ is the heat transfer coefficient between the refrigerant and evaporator or condenser shell; $A_{ei}$ is the inside evaporator area; $t_{es}$ is the temperature of the evaporator shell; $C_{er}$ and $M_{er}$ are the specific heat capacity and mass quality of the refrigerant in the evaporator; $\alpha_w$ is the heat transfer coefficient between the water and evaporator or condenser shell; $A_{eo}$ is the outside evaporator area; $t_{ewm}$ denotes the mean temperature of the chilled water in the evaporator; $C_{s}$ and $M_{es}$ represent the specific heat capacity and mass quality of the evaporator shell; $G_{ec}$ is the chilled water flow rate and one of manipulated inputs in the electric chiller; $t_{rec}$ is the aforementioned return water temperature on the chiller side; $t_{ec}$ is the chilled water temperature supplied by the electric chiller and is calculated by $t_{ec} = 2t_{ewm} - t_{rec}$; $M_{ew}$ is the mass quality of the chilled water in the evaporator; $t_c$ denotes the condensing temperature in the condenser; $h_cri$ and $h_cro$ represent the enthalpy of refrigerant at the condenser inlet and outlet, respectively; $A_{ci}$ is the inside condenser area; $t_{cs}$ is the temperature of the condenser shell; $C_{cr}$ and $M_{cr}$ are the specific heat capacity and mass quality of the refrigerant in the condenser; $A_{co}$ is the outside condenser area; $t_{ewm}$ denotes the mean temperature of cooling water in the condenser; $M_{cs}$ represents the mass quality of the condenser shell; $G_{cw}$ is the cooling water flow rate; $t_{cwi}$ and $t_{cwo}$ represent the temperature of the cooling water at at the condenser inlet and outlet; $M_{cw}$ is the mass quality of the cooling water in the condenser. For $G_r$, it can be achieved from:
\begin{equation}
	G_r = \eta_{vl}N_{ec}\rho_{eg}V_{cp}
\end{equation}
where $\eta_{vl} = 0.98 - 0.085\left(p_r^{1/k_r}-1\right)$ is the volumetric efficiency, $k_r$ is the adiabatic index of refrigerant, $p_r$ is the pressure ratio depending on $p_r = \frac{p_c}{p_e}$ in which $p_c$ and $p_e$ are the evaporation and condensing pressure evaluated by:
\begin{equation}
	p_c = \frac{e^{21.3-\frac{2025.5}{248.94+t_c}}}{10^6}
\end{equation}
\begin{equation}
	p_e = \frac{e^{21.3-\frac{2025.5}{248.94+t_e}}}{10^6}
\end{equation}
 $N_{ec}$ is compressor frequency and another manipulated input; $\rho_{eg}$ and $V_{cp}$ are the density of refrigerant and the volume of the compressor. The electric power consumption of the compressor $P_{cp}$ is computed by:
\begin{equation}
	P_{cp} = z_{ec}\frac{G_r w_i}{\eta_{cp}}
\end{equation}
where $w_i$ is the specific power of compressor evaluated by:
\begin{equation}
	w_i=\frac{k_r}{10^3(k_r-1)}\frac{10^6p_e}{\rho_{eg}}\left(p_r^{\frac{k_r-1}{k_r}}-1\right)
\end{equation}
$\eta_{cp}$ is the compressor efficiency from $\eta_{cp} = 0.9085e^{-0.06443p_r}-7.605e^{-3.155p_r}$. The cooling power of the electric chiller can also be calculated from $Q_{ec} = G_{ec}C_w(t_{rec} - t_{ec})$. Besides, we assume that the degree of superheat can be held by regulating the expansion valve \cite{r43}.

\subsubsection{Chilled water storage unit}
In the IES, the chilled water storage unit stores the chilled water and return water in the cold and hot tank, respectively \cite{r47}, \cite{r48}. According to the conservation of mass and energy, the dynamics of the chilled water storage unit are described as follows:
\begin{equation}
	\frac{d C_{sot}}{d\tau} = -\frac{G_{st}}{M_{st}} \label{26}
\end{equation}
\begin{equation}
	\frac{d C_{stc}}{d\tau} = -G_{st}t_{cp}
\end{equation}
\begin{equation}
	\frac{d C_{sth}}{d\tau} = G_{st}t_{hp}
\end{equation}
where $C_{sot}$ is the capacity state of the cold tank that indicates the ratio of the remaining cold water $m_{stc}$ to the maximum mass capacity of the cold tank $M_{st}$; therefore, $m_{stc}$ is calculated from $m_{stc} = C_{sot}M_{st}$; $G_{st}$ represents the water flow rate of the tanks, evaluated by $G_{st} = z_{st}G_{stu} + (z_{st} - 1)G_{stu}$ where $z_{st}$ is an integer variable indicating the cooling discharging or charging by 1 and 0; under cooling discharging mode, $G_{st}$ is positive, otherwise $G_{st}$ is negative; $G_{stu}$ is the absolute value of the water flow rate $G_{st}$ and taken as a manipulated variable; $C_{stc}$ is the heat capacity of cold water in the cold tank; $t_{cp}$ represents the water temperature in the pipe of the cold tank and is obtained from $t_{cp} = z_{st}t_{stc} + (1 - z_{st})t_{slc}$ where $t_{stc}$ is the temperature of cold water in the cold tank and evaluated by $t_{stc} = C_{stc}/m_{stc}$ and $t_{slc}$ is the chilled water temperature from the chillers; $C_{sth}$ is the heat capacity of hot water in the hot tank; $t_{hp}$ is the water temperature in the pipe of the hot tank and is calculated from $t_{hp} = z_{st}t_{re} + (1 - z_{st})t_{sth}$ where $t_{re}$ is the return water temperature from the customers and $t_{sth}$ is the temperature of hot water in the hot tank and evaluated by $t_{sth} = C_{sth}/m_{sth}$ and $m_{sth}$, the mass quality of remaining hot water in the hot tank, can be obtained from $m_{sth} = (1 - C_{sot})M_{st}$. The expressions of $t_{slc}$ and $t_{re}$ will be given in Section 2.1.6. The cooling power of the chilled water storage unit, $Q_{st}$, is achieved from $Q_{st} = G_{st}C_w(t_{hp} - t_{cp})$, where $Q_{st}$ is that cooling discharging if $Q_{st}$ is positive, or charging if negative.

\subsubsection{Building and auxiliary pipeline}

In the IES, the pipeline plays a part in transmitting the cooling. According to the conservation of mass and energy, the water flow rate and temperature at different locations, marked in Figure \ref{f1}, of the pipeline is expressed as follows:
\begin{equation}
	t_{rec} = \frac{G_{sl}t_{re} - G_{st}t_{hp}}{G_{ab} + G_{ec}}
\end{equation}
\begin{equation}
	t_{slc} = \frac{G_{ab}t_{ab} + G_{ec}t_{ec}}{G_{ab} + G_{ec}}
\end{equation}
\begin{equation}
	t_{sl} = \frac{G_{ab}t_{ab} + G_{ec}t_{ec} + G_{st}t_{cp}}{G_{sl}}
\end{equation}
where $t_{rec}$ is the return water temperature on the chiller side; $G_{sl}$ is the flow rate of the supply water for customers evaluated by $G_{sl} = G_{ab} + G_{ec} + G_{st}$; $t_{re}$ is the return water temperature from customers and will be further discussed later; $t_{slc}$ represents the temperature of chiller water from the absorption and electric chiller; $t_{sl}$ is the final supply water temperature. And besides, the electric power of the pump $P_{pmp}$ is presents as follows:
\begin{equation}
	P_{pmp} = \frac{G_{all} g_e H_{pmp}}{10^3\eta_{pmp}}
\end{equation}
where $G_{all}$ represents the total flow in the pipeline evaluated via $G_{all} = G_{ab} + G_{ec} + G_{stu}$; $g_e$ is the gravitational acceleration; $H_{pmp}$ is the hydraulic head depending on $H_{pmp} = S_{pmp}G^2_{all}$ \cite{r44} where $S_{pmp}$ is the resistance coefficient; $\eta_{pmp}$ is the pump performance curve $\eta_{pmp} = b_{pmp,1}G_{all}^2 + b_{pmp,2}G_{all} + b_{pmp,3}$ \cite{r45}, and $b_{pmp,i},\ i=1,2,3$ being the coefficients.

As mentioned before, the supply water, the mix of chilled water from the chillers and cold water from the cold tank, will transfer heat with air in the fan coil units to cool the building. In this IES, a performance curve for fan coil units, modeled in \cite{r37}, is employed and written as follows:
\begin{equation}
	Q_{fcu} = Q_{fcu,0}\frac{t_{br} - t_{sl}} {\Delta t_{wa,0}} \left(\frac{G_{sl}}{G_{sl,0}}\right)^{0.6}
\end{equation}
where $Q_{fcu}$ is the calculated cooling power in the fan coil units; $Q_{fcu,0}$ denote the cooling power under nominal conditions; $t_{br}$ is the temperature of building; $\Delta t_{wa,0}$ and $G_{sl,0}$ represent the nominal temperature difference between the building air and supply water and the nominal supply water flow rate, respectively. Then the calculated temperature of the return water, $t_{re,cal}$, can be obtained by $t_{re,cal} = t_{sl} + Q_{fcu}/(G_{sl}C_w)$. Furthermore, a first-order inertial is used to capture the thermal inertial of the fan coil units as follows:
\begin{equation}
	\frac{d t_{re}}{d\tau} = \frac{t_{re,cal} - t_{re}}{\tau_{fcu}} \label{34}
\end{equation}
Consequently, the total cooling power providing for customers is formulated as $Q_{sl} = G_{sl}C_w(t_{re} - t_{sl})$. The dynamics of the temperature of building, $t_{br}$, is captured by following differential equation \cite{r16}:
\begin{equation}
	\frac{d t_{br}}{d\tau} = \frac{U_{br}(t_a - t_{br}) - Q_{sl} + Q_{other}}{C_{br}} \label{35}
\end{equation}
where $U_{br}$ is the scaled heat transfer coefficient of building; $t_a$ is the same ambient temperature as described in the photovoltaic energy generation unit; $Q_{other}$ represents the other cooling load; $C_{br}$ is the heat capacity of the building.

\subsubsection{Battery bank}
As aforementioned, the battery bank will fill the gap in electric power in the IES through a battery management system \cite{r57}. In this IES, a lithium-ion battery bank based on the Thevenin equivalent model \cite{r38} is introduced as follows:
\begin{equation}
	\frac{d v_{cap}}{d\tau} = \frac{i_{ba}}{C_{1b}} - \frac{v_{cap}}{R_{1b}C_{1b}} \label{36}
\end{equation}
\begin{equation}
	\frac{d C_{soc}}{d\tau} = \frac{-i_{ba}}{3600C_{eb}} \label{37}
\end{equation}
\begin{equation}
	V_{ba} = n_{sb}(E_m - v_{cap} - R_{0b}i_{ba})
\end{equation}
where $v_{cap}$ is the capacitor voltage; $i_{ba}$ is the current in each battery cell and its expression will be given later; $C_{1b}$ and $R_{1b}$ represent the capacitance and series resistance in the battery cells; $C_{soc}$ is the state of charge, also called the capacity state of battery, which reflects the proportion of the left charge in the battery bank; $C_{eb}$ is the total charge in the battery cells; $V_{ba}$ denotes the voltage of the battery bank; $n_{sb}$ represents the number of battery cells in series; $E_m$ is the total charge and electric potential in each battery cell; $R_{0b}$ is the parallel resistance. For $i_{ba}$, it cab be computed by $i_{ba} = I_{ba}/n_{pb}$ where $n_{pb}$ is the number of battery cells in parallel, $I_{ba}$ represents the current through the battery bank evaluated by:
\begin{equation}
	\frac{d I_{ba}}{d\tau} = \frac{I_{ba,cal} - I_{ba}}{\tau_{dc}} \label{39}
\end{equation}
where $I_{ba,cal}$ is the calculated current by $I_{ba,cal} = 10^3P_{ba,d}/V_{ba}$, in which $P_{ba,d}$ is the electric power needed to be supplied by the battery bank and is evaluated by $P_{ba,d} = P_d + P_{cp} + P_{pmp} - P_{pv} - P_{fc} - P_{mt}$; $P_d$ is customers' demands for electric power; $\tau_{dc}$ indicates the time constant of the regulation. Then the electric power of the battery bank $P_{ba}$ is presented as follows:
\begin{equation}
	P_{ba} = \frac{V_{ba}I_{ba}}{10^3}
\end{equation}
When $P_{ba}$ is positive, the battery bank is discharged, otherwise, the battery bank is charged.

Finally, the total electric power provided by the IES to the building can be evaluated as follows:
\begin{equation}
	P_{sl} = P_{pv} + P_{fc} + P_{mt} + P_{ba} - P_{cp} - P_{pmp} \label{eq_add_2}
\end{equation}

The main parameters used in the stand-alone IES Eqs.\eqref{eq_add_1}-\eqref{eq_add_2} are displayed in Table \ref{t2}.
\begin{table}[t] \small 
	\centering
	\caption{Main model parameters}
	\label{t2}
	\renewcommand{\arraystretch}{1}
	\tabcolsep 0.1pt
	\begin{tabular}{llllll} \hline
		Parameter & Value & Parameter & Value & Parameter & Value\\ \hline
		$n_{pp}$& 35 & $\tau_{abw}$ (s)& 80& $t_{cwi}$ ($^{\circ}$C)& 28\\
		$n_{sp}$& 6 & $\tau_{abt}$ (s)& 70& $E_m$ (V)& 50\\
		$c_{pv,1}$ (1/$^{\circ}$C)& 2.5$\times 10^{-3}$& $C_w$ (kJ/$^{\circ}$C)& 4.2& $R_{1b}$ ($\Omega$)& 4.8$\times 10^{-3}$\\
		$c_{pv,2}$ (1/(W/m$^2$))& 9.2$\times 10^{-5}$& $k_{ma,1}$& -6.862$\times 10^5$& $R_{0b}$ ($\Omega$)& 4.2$\times 10^{-2}$\\
		$c_{pv,3}$ (1/$^{\circ}$C)& 4.13$\times 10^{-3}$& $k_{ma,2}$& 2.144$\times 10^{4}$& $C_{1b}$ (F)& 1296\\
		$I_{max,0}$ (A)& 7.412& $k_{ma,3}$& -1.128$\times 10^{2}$& $n_{pb}$& 4\\
		$V_{max,0}$ (V)& 29& $k_{ma,4}$& -3.863$\times 10^{-2}$& $n_{sb}$& 4\\
		$\tau_e$ (s)& 0.8& $k_{ma,5}$& 1.24& $C_{eb}$ (C)& 200\\
		$\tau_f$ (s)& 5& $k_{ma,6}$& -4.429& $\tau_{dc}$ (s)& 0.8\\
		$\tau_{O_2}$ (s)& 2.91& $k_{ma,7}$& 0.957& $M_{st}$ (kg)& 2$\times 10^{4}$\\
		$\tau_{H_2O}$ (s)& 78.3 & $k_{ma,8}$& -11.484& $S_{pmp}$ ((m$\cdot$s$^2$)/kg$^2$)& 5.204\\
		$\tau_{H_2}$ (s)& 26.1 & $A_{ci}$ (m$^2$)& 2.477& $b_{pmp,1}$& -1.007$\times 10^{-2}$\\
		$K_{O_2}$ (mol/(atm$\cdot$s))& 2.52 & $A_{co}$ (m$^2$)& 2.831& $b_{pmp,2}$& 1.308$\times 10^{-1}$\\
		$K_{H_2O}$ (mol/(atm$\cdot$s))& 0.281 & $A_{ei}$ (m$^2$)& 2.638& $b_{pmp,3}$& 3.552$\times 10^{-1}$\\
		$K_{H_2}$ (mol/(atm$\cdot$s))& 0.843 & $A_{eo}$ (m$^2$)& 3.015& $Q_{fcu,0}$ (kW)& 125\\
		$r_{fc}$ ($\Omega$)& 0.126& $\alpha_r$ (kW/(m$^2$ $\cdot ^{\circ}$C))& 4.476& $\Delta t_{wa,0}$ ($^{\circ}$C)& 15\\
		$a_{fc}$& 0.05& $\alpha_w$ (kW/(m$^2$ $\cdot ^{\circ}$C))& 5.08& $G_{sl,0}$ (kg/s)& 5.952\\
		$b_{fc}$& 0.11& $C_{cr}$ (kJ/(kg $\cdot ^{\circ}$C))& 1.339& $\tau_{fcu}$ (s)& 20\\
		$R_{0fc}$ (J/(mol$\cdot$K))& 8.314& $C_{er}$ (kJ/(kg $\cdot ^{\circ}$C))& 1.175& $C_{br}$ (kJ/$^{\circ}$C)& 5.14$\times 10^{4}$\\
		$T_{0fc}$ (K)& 1273& $C_s$ (kJ/(kg $\cdot ^{\circ}$C))& 0.394& $U_{br}$ (kW/$^{\circ}$C)& 4.063\\
		$F_{0fc}$ (C/mol)& 96485& $k_r$& 1.388& $k_{fc}$& 8.889$\times 10^{3}$\\
		$I_L$ (A)& 800& $M_{cr}$ (kg)& 9.784& $k_{mt}$& 1.1956 $\times 10^{4}$\\
		$K_r$ (mol/(A$\cdot$s))& 0.996$\times 10^{-3}$& $M_{cs}$ (kg)& 94.487& $k_{ab}$& 0.9375\\
		$N_{0fc}$ & 384& $M_{cw}$ (kg)& 79.762& $k_{ec}$& 0.4546\\
		$E_{0fc}$ (V)& 1.18& $M_{er}$ (kg)& 11.768& $k_{cp}$& 0.252\\
		$r_{H-O}$& 1.145& $M_{es}$ (kg)& 100.61& $k_{st}$& 21\\
		$P_{mt,0}$ (kW)& 80& $M_{ew}$ (kg)& 71.396& $k_{pmp}$& 2.3352\\
		$t_{ab,0}$ ($^{\circ}$C)& 7& $\rho_{eg}$ (kg/m$^3$)& 22.602& $k_{ba}$& -1.7361$\times 10^{-6}$\\
		$\tau_{mtf}$ (s)& 20& $V_{cp}$ (m$^3$)& 1.422$\times 10^{-4}$& $k_{pv}$& 4.5139$\times 10^{-2}$\\
		$\tau_{abf}$ (s)& 130& $G_{cw}$ (kg/s)& 2.832& $G_{ab,0}$ (kg/s)& 3.5714\\
		$S_{ra,0}$ (W/m$^2$)& 1000& $t_{pv,0}$ ($^{\circ}$C)& 25& $G_{ec,0}$ (kg/s)& 2.381\\
		$M_{ng}$ (1/(kg/mol))& 62.5& $fu_d$& 0.8& $g_e$ (m/s$^2$)& 9.8\\ \hline
	\end{tabular}
\end{table}

\subsection{Control problem formulation}
For the stand-alone IES, two important output variables are the electric power supplied to the customers ($P_{sl}$), the building temperature ($t_{br}$). Six continuous manipulated variables are the mass flow rates of natural gas in the fuel cell and microturbine ($G_{ff}$ and $G_{fm}$), the chilled water flow rate in the absorption chiller ($G_{ab}$), the compressor frequency in the electric chiller ($N_{ec}$), the chilled water flow rate in the electric chiller ($G_{ec}$), and the absolute value of the water flow rate in the tanks ($G_{stu}$). Four 0-1 integer manipulated variables are the start-up/shut-down of the fuel cell ($z_{fc}$), the start-up/shut-down of the microturbine integrated with absorption chiller ($z_{ma}$), the start-up/shut-down of the electric chiller ($z_{ec}$), and the cooling charging/discharging of the chilled water storage unit ($z_{st}$). Four general disturbances are the ambient temperature ($t_a$), the solar radiation ($S_{ra}$), the electric power demand ($P_d$), and the other cooling load ($Q_{other}$). Let us define the state vector as $x = [I_f, G_{H_{2}}, p_{O_2}, p_{H_{2}O}, p_{H_{2}}, P_{mtf}, t_{abf}, t_{abw}, t_{abt}, t_c, t_{cs}, t_{cwm}, t_e, t_{es}, t_{ewm}, v_{cap}, C_{soc}, I_{ba}, C_{sot}, C_{stc}, C_{sth},\\ t_{re}, t_{br}]^T$, the continuous manipulated inputs as $u = [G_{ff}, G_{fm}, G_{ab}, N_{ec}, G_{ec}, G_{stu}]^T$, the integer manipulated inputs as $z = [z_{fc}, z_{ma}, z_{ec}, z_{st}]^T$, the disturbances vector as $\omega = [t_a, S_{ra}, P_d, Q_{other}]^T$, and the process output vector as $y = [P_{sl}, t_{br}]^T$. Then the stand-alone IES can be described by a compact nonlinear form as follows:
\begin{equation}
	\begin{aligned}
		&\dot{x}(\tau)=f(x(\tau),u(\tau), z(\tau), \omega (\tau))\\
		&y(\tau)=h(x(\tau),u(\tau), z(\tau), \omega (\tau)) \label{42}
	\end{aligned}
\end{equation}
where $x\in \mathbb{R}^{23}$, $u\in \mathbb{R}^{6}$, $z\in \{0,1\}^4$, $\omega \in \mathbb{R}^{4}$ and $y\in \mathbb{R}^{2}$.

The stand-alone IES is a large-scale complex thermodynamic and electric nonlinear system. Several factors make the reliable and economic operation of this stand-alone IES challenging. First, there is time-scale multiplicity in the dynamics of the IES, which can lead to ill-conditioned optimization or even erroneous behavior in the close-loop system and a significant computational burden if a typical centralized controller is used without explicitly considering the time-scale multiplicity. Second, the stand-alone IES is directly exposed to environmental conditions and customers' needs. The optimal operating point of the system needs to vary promptly according to external conditions to satisfy the customers' demands for electricity and cooling while reducing fuel consumption. Meanwhile, it is worth noting that customers and IESs generally pay more attention on tracking electric power reference than building temperature due to microgrids' safety, which also intensifies the operational challenges caused by the couplings in IESs. Furthermore, environmental conditions and customers' demands usually change periodically every day in a long-term view. Therefore, the start-ups/shut-downs of the operating units and the operational curves of energy storage units should be consistent with long-term environmental conditions and customers' demands forecasts. Three typical performance indices for the operation of the stand-alone IES are shown below:
\begin{equation}
	J_1=\|y_1-y_{sp,1}\|^2 \label{a45}
\end{equation}
\begin{equation}
	J_2=\|y_2-y_{sp,2}\|^2 \label{a46}
\end{equation}
\begin{equation}
	J_3= u_1+u_2 \label{a47}
\end{equation}
where $y_{sp,1}$ is the customers' demand for the electric power, $y_{sp,2}$ is the demand for building temperature, $J_1$ is the electric power tracking error, $J_2$ is the building temperature tracking error, $J_3$ is the total consumption of natural gas. These performance indices will be taken into account in the proposed composite EMPC to optimize and coordinate system operation.

\section{Multi-time-scale subsystem decomposition}
As aforementioned, the stand-alone IES is essentially a large-scale and complex nonlinear system exhibiting time-scale multiplicity in its dynamics. The IES is inherently associated with time-scale multiplicity due to the strong coupling of physical processes and the presence of obviously different time constants and residence times in various operating units. The centralized control schemes are no longer suitable for the IES. In this case, it is a practical solution to decompose the IES described in Eq.\eqref{42} into several separate reduced-order subsystems according to different time scales to facilitate the design of a control scheme. Therefore, in this section, the time-scale multiplicity in the IES is first investigated and identified as the slow, medium and fast time scales. Then the entire IES is decomposed into reduced-order slow, medium, and fast subsystems according to these time scales. The reduced-order subsystems will be used to match the dynamic time-scale multiplicity in the design of the proposed composite EMPC scheme later.

\subsection{Investigation of the IES time-scale multiplicity}
First, to perform the multi-time-scale decomposition, it is necessary to identify and compare the time constants of each of the states. By analyzing the dynamic characteristics of the IES, we define the following parameters to represent the approximate time constants or residence times for each state in Eq.\eqref{42}: 
\begin{equation}
	\begin{aligned}
		&\tau_{x,1} = \tau_e,\ \tau_{x,2} = \tau_f,\ \tau_{x,3} = \tau_{O_2},\ \tau_{x,4} = \tau_{H_2O},\ \tau_{x,5} = \tau_{H_2} \\
		&\tau_{x,6} = \tau_{mtf},\ \tau_{x,7} = \tau_{abf},\ \tau_{x,8} = \tau_{abw},\ \tau_{x,9} = \tau_{abt} \\
		&\tau_{x,10} = \frac{C_{cr}M_{cr}}{\alpha_r A_{ci}},\ \tau_{x,11} = \frac{C_s M_{cs}}{\alpha_r A_{ci} + \alpha_w A_{co}},\ \tau_{x,12} = \frac{C_w M_{cw}}{\alpha_w A_{co}} \\
		&\tau_{x,13} = \frac{C_{er}M_{er}}{\alpha_r A_{ei}},\ \tau_{x,14} = \frac{C_s M_{es}}{\alpha_r A_{ei} + \alpha_w A_{eo}},\ \tau_{x,15} = \frac{C_w M_{ew}}{\alpha_w A_{eo}} \\
		&\tau_{x,16} = R_{1b}C_{1b},\ \tau_{x,17} = \frac{3600C_{eb}\beta_{ba}}{i_{ba}},\ \tau_{x,18} = \tau_{dc}\\ &\tau_{x,19} = \tau_{x,20} = \tau_{x,21} = \frac{M_{st}\beta_{st}}{G_{st}},\ \tau_{x,22} = \tau_{fcu},\ \tau_{x,23} = \frac{C_{br}}{U_{br}} \label{62}
	\end{aligned}
\end{equation}
where $\tau_{x,i}, \ i=1,\dots,23$ denote the approximate time constant or residence time of the $i$-th state in Eq.\eqref{42}, and $\beta_{ba} = 0.8$ and $\beta_{st} = 0.9$ are the limitation to the available capacity of the battery and cold tank respectively. By comparing the magnitudes of these different time constants in Eq.\eqref{62}, they are divided into three sets as follows:
\begin{equation}
	\begin{aligned}
		&\mathbb{T}_s = \{\tau_{x,17},\tau_{x,19},\tau_{x,20},\tau_{x,21},\tau_{x,23}\} \\
		&\mathbb{T}_m = \{\tau_{x,4},\tau_{x,5},\tau_{x,6},\tau_{x,7},\tau_{x,8},\tau_{x,9},\tau_{x,12},\tau_{x,15},\tau_{x,22}\}\\
		&\mathbb{T}_f = \{\tau_{x,1},\tau_{x,2},\tau_{x,3},\tau_{x,10},\tau_{x,11},\tau_{x,13},\tau_{x,14},\tau_{x,16},\tau_{x,18}\} \label{63}
	\end{aligned}
\end{equation}
where the elements in $\mathbb{T}_s$ have significantly lager values ($12652.31s\sim18000s$) compared to the elements in $\mathbb{T}_m$ ($20s\sim130s$) and $\mathbb{T}_f$ ($0.8s\sim6.221s$), while the values of the elements in $\mathbb{T}_f$ are considerably smaller than those in $\mathbb{T}_s$ and $\mathbb{T}_m$. This suggests that the states corresponding to the elements in $\mathbb{T}_s$ have a slow time-scale, the states corresponding to the elements in $\mathbb{T}_f$ exhibit significantly faster dynamics. The time scales of the corresponding states in $\mathbb{T}_m$ are between in $\mathbb{T}_s$ and $\mathbb{T}_f$.

Next, based on the above classification of the states, we proceed to develop the reduced-order models using the singular perturbation theory. The singular perturbation theory requires the identification of a dimensionless small parameter, $\epsilon$ $(0\textless\ \epsilon \ll 1)$ that indicates the existence of the time-scale separation in the system, to decompose the system \cite{r30}, \cite{r49}. To this end, $\tau_{x,23}$, $\tau_{x,9}$, $\tau_{x,2}$ are selected as the representative time constant of the corresponding set, respectively. Specifically, $\tau_{x,23} = 12652.31 s$, $\tau_{x,9} = 70 s$, $\tau_{x,2} = 5 s$. We note that $\tau_{x,23}$, $\tau_{x,9}$, and $\tau_{x,2}$ exhibit significantly different values. Based on these representative values, we define the following dimensionless small parameters:
\begin{equation}
	\epsilon_1 = \frac{\tau_{x,2}}{\tau_{x,23}}=0.000395\ll 1,\ \epsilon_2 = \frac{\tau_{x,9}}{\tau_{x,23}}=0.00553\ll 1,\ \epsilon_3 = \frac{\epsilon_1}{\epsilon_2}=0.0714\ll 1
\end{equation}
where $\epsilon_3$ indicates that $\epsilon_1$ is an infinitesimal of higher order than $\epsilon_2$.

By taking these into account, the IES model of Eq.\eqref{42} can be rewritten as the following singular perturbation form by dividing Eqs.\eqref{4}-\eqref{8}, \eqref{11}-\eqref{14}, \eqref{17}-\eqref{22}, \eqref{34}, \eqref{36} and \eqref{39} by $\tau_{x,23}$:
\begin{subequations} \label{65}
	\begin{align}
		\frac{d x_s}{d\tau} &= f_s(x_s,x_m,u_s,z,\omega) \\
		\epsilon_2 \frac{d x_m}{d\tau} &= f_m(x_s,x_m,x_f,u_s,u_m,z) \\
		\epsilon_1 \frac{d x_f}{d\tau} &= f_f(x_m,x_f,u_s,u_f,z,\omega) \\
		y_s(\tau) &= h_s(x_s) \label{65d} \\
		y_{mf}(\tau) &= h_{mf}(x_m,x_f,u_s,u_f,z,\omega) \label{65e}
	\end{align}
\end{subequations}
where $x_s = [x_{17}, x_{19}, x_{20}, x_{21}, x_{23}]^T$ represents the slow dynamics in the original IES Eq.\eqref{42} and consists of the states in the building and chilled water storage unit and part of the states in the battery bank; $x_m$ is the medium dynamics in Eq.\eqref{42} and defined as $x_m = [x_4, x_5, x_6, x_7, x_8, x_9, x_{12},\\ x_{15}, x_{22}]^T$ in which the states in the microturbine integrated with absorption chiller and part of the states in the fuel cell and electric chiller are contained; $x_f = [x_1, x_2, x_3, x_{10}, x_{11}, x_{13}, x_{14}, x_{16}, x_{18}]^T$ indicates the fast dynamics in Eq.\eqref{42} which comprises the rest of states in the battery bank, fuel cell and electric chiller.

According to the physical connection between the continuous manipulated inputs and the states of the IES, the continuous manipulated inputs are separated into three sets correspondingly as follows: $u_s = [u_3, u_5, u_6]^T$ explicitly affects the slow, medium and fast process states $x_s$, $x_m$, $x_f$, and will be determined by the slow EMPC introduced later, and $u_m = u_2$ directly correlates to the medium states $x_m$ and will be regulated by the medium EMPC, $u_f =[u_1, u_4]^T$ explicitly connects the fast states $x_f$ and will be determined by the fast EMPC. For the integer manipulated inputs $z$, it keeps the original form of $z = [z_1, z_2, z_3, z_4]^T$ that will be optimized by the long-term EMPC. With respect to the controlled output decomposition, $y_s = y_2$, the building temperature, just directly depends on the slow states, while $y_{mf} = y_1$, the total electric power provided by the IES, depends on the medium and fast states and the manipulated inputs $u_s$ and $u_f$. In this case, the performance index $J_1$ in Eq.\eqref{a45}, which is the electricity tracking error, is correlated with the slow, medium and fast subsystem; the performance index $J_2$ in Eq.\eqref{a46}, representing the building temperature tacking error is just associated with the slow subsystem; the performance index $J_3$ in Eq.\eqref{a47} indicating the fuel consumption is relevant to the medium and fast subsystem.

\subsection{The reduced-order subsystem based on multi-time-scale decomposition}

Based on the investigation of the time-scale multiplicity in the IES dynamics, we can move forward with multi-time-scale separation for the subsystem decomposition. To derive the reduced-order slow subsystem, we set $\epsilon_1,\epsilon_2 \rightarrow 0$ in Eq.\eqref{65}, and obtain the expression of the reduced-order slow subsystem as follows:
\begin{subequations} \label{66}
	\begin{align}
		\frac{d x_s}{d\tau} &= f_s(x_s,x_m,u_s,z,\omega) \label{66a}\\
		0 &= f_m(x_s,x_m,x_f,u_s,u_m,z) \label{66b}\\
		0 &= f_f(x_m,x_f,u_s,u_f,z,\omega) \label{66c}
	\end{align}
\end{subequations}
We assume that a unique solution exists in the algebraic equations \eqref{66b} and \eqref{66c}, which indicates the pair of $(x_m,x_f)$ can be uniquely expressed in terms of $x_s$, $u_s$, $u_m$, $u_f$, $z$ and $\omega$. This assumption is a standard formulation in time-scale based decomposition and does not impose restrictions on practical control system for the IES. Consequently, the Eqs.\eqref{66a}-\eqref{66c} and Eqs.\eqref{65d} and \eqref{65e} are rewritten as a more compact form as follows:
\begin{subequations} \label{67}
	\begin{align}
		\frac{d x_s}{d\tau} &= f_{sc}(x_s,u_s,u_m,u_f,z,\omega) \label{67a}\\
		y_s(\tau) &= h_s(x_s) \label{67b}\\
		y_{mf}(\tau) &= h^s_{mf}(x_s,u_s,u_m,u_f,z,\omega) \label{67c}
	\end{align}
\end{subequations}

Subsequently, to derive the reduced-order medium subsystem from Eq.\eqref{65}, we define a medium (stretched) time scale $\tau_m = \frac{\tau}{\epsilon_2}$ and substitute $\tau_m$ into Eq.\eqref{65} to obtain that: 
\begin{subequations}
	\begin{align}
		\frac{d x_s}{d\tau_m} &= \epsilon_2 f_s(x_s,x_m,u_s,z,\omega) \label{add_e1a}\\
		\frac{d x_m}{d\tau_m} &= f_m(x_s,x_m,x_f,u_s,u_m,z) \label{add_e1b}\\
		\epsilon_3 \frac{d x_f}{d\tau_m} &= f_f(x_m,x_f,u_s,u_f,z,\omega) \label{add_e1c}
	\end{align}
\end{subequations}
Then setting $\epsilon_2 \rightarrow 0$ (which also suggests $\epsilon_3 \rightarrow 0$ because of $\epsilon_1 \ll \epsilon_2$), the reduced-order medium subsystem is achieved as follows: 
\begin{subequations}
	\begin{align}
		\frac{d x_s}{d\tau_m} &= 0 \label{68a}\\
		\frac{d x_m}{d\tau_m} &= f_m(x_s,x_m,x_f,u_s,u_m,z) \label{68b}\\
		0 &= f_f(x_m,x_f,u_s,u_f,z,\omega) \label{68c}
	\end{align}
\end{subequations}
Eq.\eqref{68a} implies that the slow dynamics $x_s$ can be considered as a constant in the medium subsystem. It is also assumed that a unique solution exists in the algebraic equation \eqref{68c}, which indicates $x_f$ can be uniquely expressed in terms of $x_m$, $u_s$, $u_f$, $z$ and $\omega$. Then the Eqs.\eqref{68a}-\eqref{68c} and Eq.\eqref{65e} can be described by a concise form as follows:
\begin{subequations} \label{69}
	\begin{align}
		\frac{d x_m}{d\tau_m} &= f_{mc}(x_s,x_m,u_s,u_m,u_f,z) \label{69a}\\
		y_{mf}(\tau_m) &= h^m_{mf}(x_m,u_s,u_f,z,\omega) \label{69b}
	\end{align}
\end{subequations}

Similarly, for deriving the reduced-order fast subsystem from Eq.\eqref{65}, we define a fast (stretched) time scale $\tau_f = \frac{\tau}{\epsilon_1}$, substitute $\tau_f$ into Eq.\eqref{65} and then set $\epsilon_1 \rightarrow 0$ (which also suggests $\epsilon_3 \rightarrow 0$), the reduced-order fast subsystem is obtained as follows: 
\begin{subequations}
	\begin{align}
		\frac{d x_s}{d\tau_f} &= 0 \label{70a}\\
		\frac{d x_m}{d\tau_f} &= 0\label{70b}\\
		\frac{d x_f}{d\tau_f} &= f_f(x_m,x_f,u_s,u_f,z,\omega) \label{70c}
	\end{align}
\end{subequations}
Further, the Eqs.\eqref{70a}-\eqref{70c} and Eq.\eqref{65e} can also be rewritten as the following expression:
\begin{subequations} \label{71}
	\begin{align}
		\frac{d x_f}{d\tau_f} &= f_{fc}(x_m,x_f,u_s,u_f,z,\omega) \label{71a}\\
		y_{mf}(\tau_f) &= h^f_{mf}(x_m,x_f,u_s,u_f,z,\omega) \label{71b}
	\end{align}
\end{subequations}
It should be noted that the slow dynamics $x_s$ and the medium dynamics $x_m$ are considered as constants in the fast subsystem Eq.\eqref{71} according to Eqs.\eqref{70a} and \eqref{70b}. In addition, both Eq.\eqref{69b} in the medium subsystem and Eq.\eqref{71b} in the fast subsystem represent the same controlled output $y_1$, the total electric power provided by the IES. But in Eq.\eqref{69b}, the evolution of the medium dynamics $x_m$ is preserved and evaluated by the differential equation \eqref{69a}, and the fast dynamics $x_f$ is expressed by $x_m$, $u_s$, $u_f$, $z$ and $\omega$ in the light of Eq.\eqref{68c}, while the Eq.\eqref{71b} take into account the evolution of $x_f$ by the differential equation \eqref{71a} and takes $x_m$ as a constant.

\begin{remark}
	It is worth noting that the multi-time-scale subsystem decomposition developed for IESs differs from those employed in existing work \cite{r56,r28,r29,r30,r49} in which the process systems are decomposed into independent slow and fast subsystems by considering two time-scale separation. In this work, the entire IES is decomposed into three different subsystems according to the time-scale multiplicity in the IES dynamics, and the necessary interconnections among these subsystems are preserved, providing a more general approach when we address the time-scale multiplicity in distributed energy systems.
\end{remark}

\section{Proposed composite economic MPC}

\begin{figure}[t]
	\centering
	\includegraphics[width=0.82\hsize]{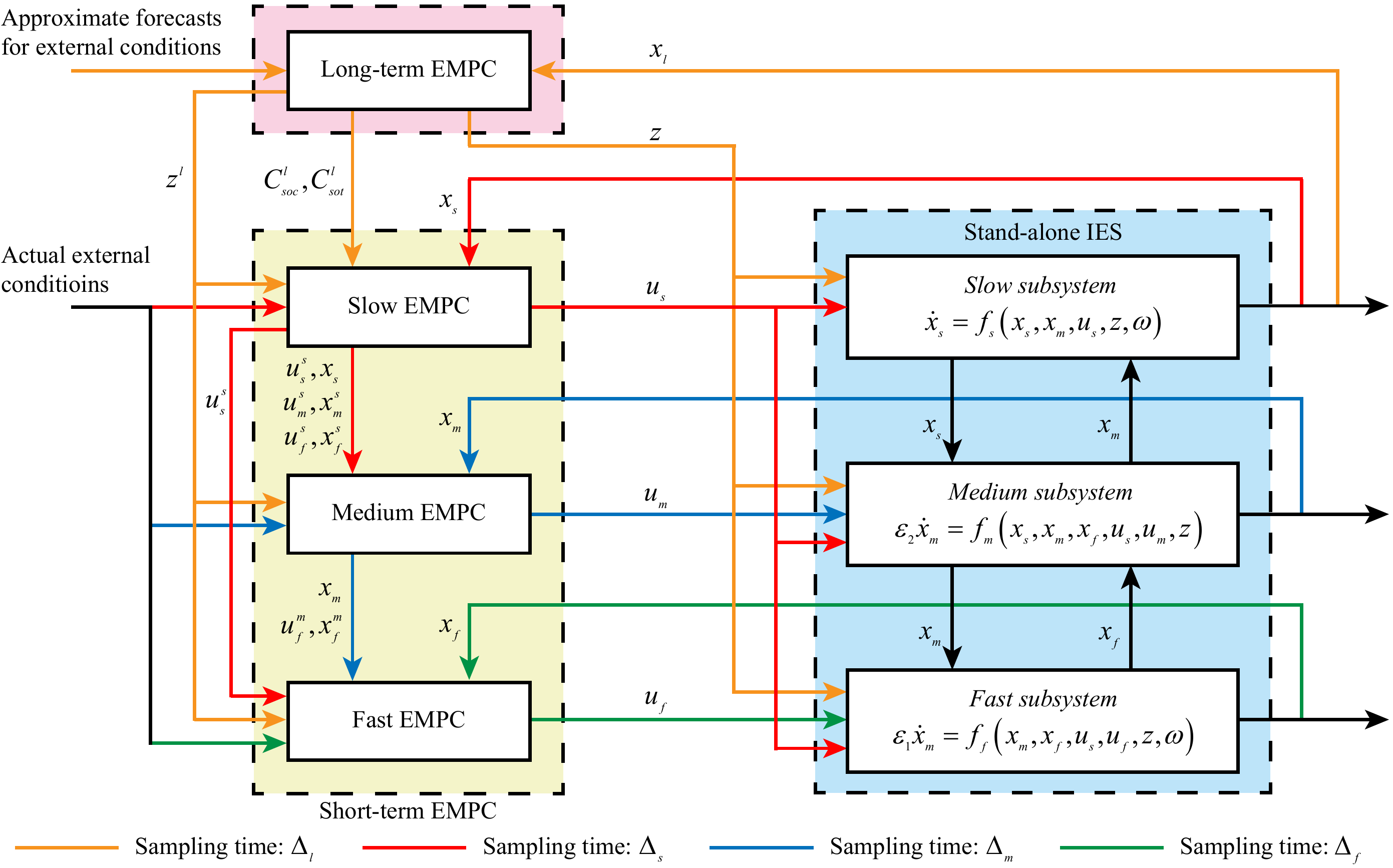}
	\caption{A schematic of proposed composite EMPC for the stand-alone IES}
	\label{f2}
\end{figure}

The IES investigated above exhibits multi-time-scale properties that make it more attractive and effective to have multiple interconnected controllers to coordinate system operation. In order to take into account the performance indices in Section 2.2 for the IES reliable and economic operation and overcome the control difficulties caused by the time-scale multiplicity, couplings, and computational complexity, this section develops a composite economic MPC with zone tracking for the stand-alone IES based on the subsystem decomposition as shown in Figure \ref{f2}. The proposed CEMPC is comprised of a long-term EMPC and a short-term EMPC that consists of three interconnected EMPCs with different time scales. These controllers are designed based on different subsystems and tasks individually.

Specifically, the long-term EMPC is proposed to ensure the system's energy balance, improve its economics, and avoid inappropriately and frequently switching operating units on/off and charging/discharging the energy storage units during a long time horizon. In the long-term EMPC, a simplified slow subsystem model derived from the slow subsystem is employed to optimize the overall behavior of the IES in 24 hours according to approximate forecasts for the external conditions and part of the slow dynamics $x_l$. From the long-term EMPC, the CEMPC will obtain the optimal capacity states of the energy storage units, $C^l_{soc}$ and $C^l_{sot}$, and the optimal integer manipulated inputs $z^l$ (where $z=z^l$) that represents the optimal decisions on the start-ups/shut-downs of the operating units and the cooling charging/discharging of the chilled water storage unit. Then $C^l_{soc}$ and $C^l_{sot}$ are sent to the short-term EMPC as references to regulate the energy storage units, while $z^l$ is applied to the IES and also sent to the short-term CEMPC. The other optimized variables are discarded, which will be re-optimized in the short-term EMPC by more accurate external conditions and system models. Then the integer variable $z$ in the short-term EMPC will be considered a known constant. The long-term EMPC will be implemented every hour. 

\begin{figure}[t]
	\centering
	\includegraphics[width=0.84\hsize]{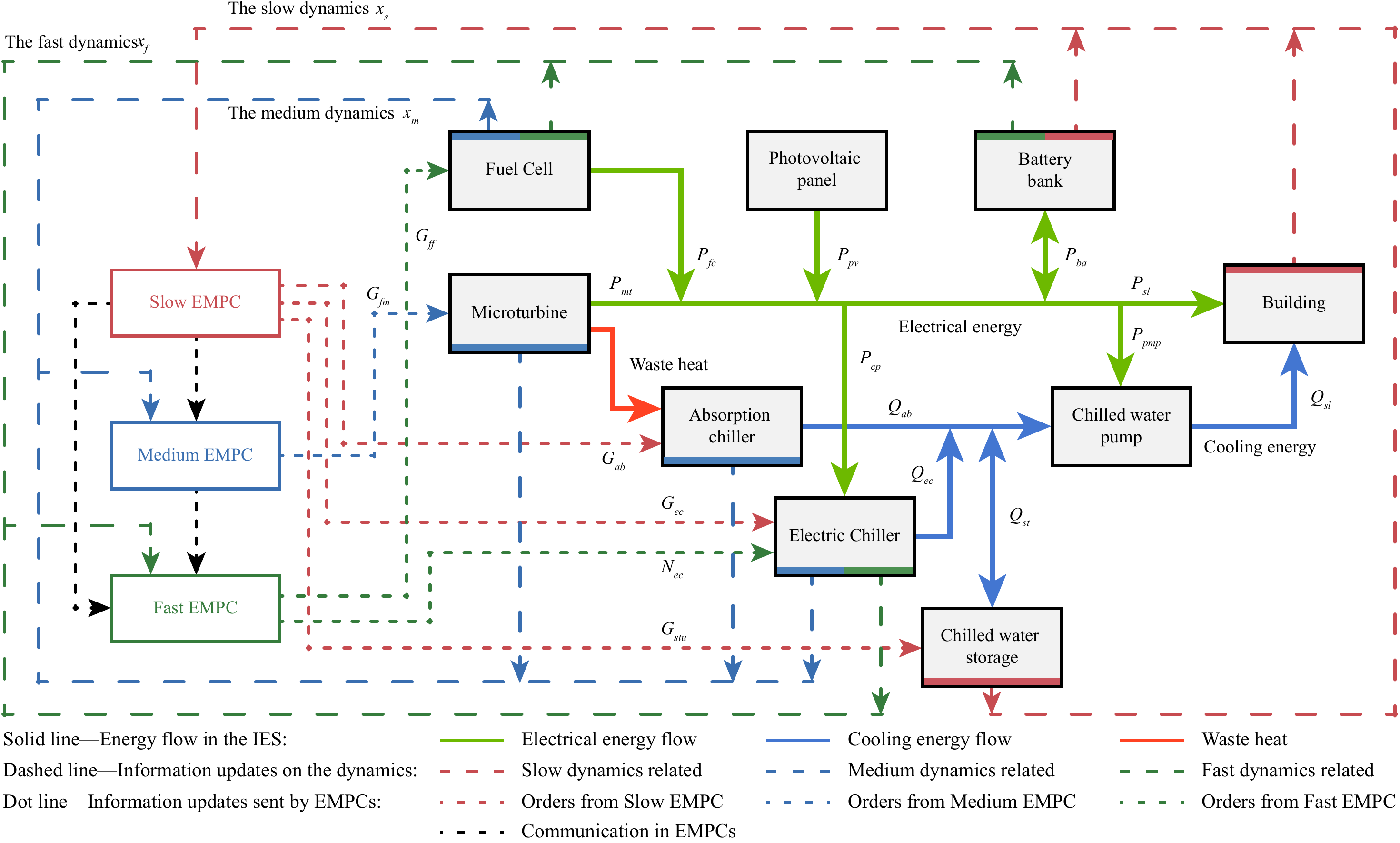}
	\caption{The information and energy flow chart in the closed-loop short-term EMPC}
	\label{add_f1}
\end{figure}

Meanwhile, the short-term EMPC is developed to precisely coordinate the IES to satisfy customers' requirements according to actual external conditions in real-time and maximize system profitability. By taking advantage of three reduced-order subsystems, the short-term EMPC consists of three interconnected EMPCs---slow, medium, and fast EMPCs. The information and energy flow chart in the closed-loop short-term EMPC are shown in Figure \ref{add_f1}. First, the slow EMPC is designed to optimize the global performance indices, $J_1$, $J_2$, and $J_3$, during dozens of minutes based on the slow subsystem and the real feedback of the slow dynamics $x_s$. The optimal $C^l_{soc}$ and $C^l_{sot}$ from the long-term EMPC are taken into consideration here. Through the slow EMPC, all of the optimal continuous manipulated inputs and states will be evaluated, in which the slow inputs $u_s$ (where $u_s=u^s_s$) will be applied to the system to regulate the operating units correlated with the slow dynamics. All required information will be sent into medium and fast EMPC simultaneously. Subsequently, the medium EMPC based on the medium subsystem is developed to regulate the medium dynamics $x_m$ by optimizing the performance indices $J_1$ and $J_3$ for dozens of seconds. In the meantime, it receives the feedback of the medium dynamics $x_m$. The optimal medium and fast states and inputs from the slow EMPC are also considered in the medium EMPC to ensure the results are consistent with the slow EMPC. Then the optimal medium and fast states and inputs will be re-optimized and achieved, in which the optimal medium inputs $u_m$ will enter the system. The necessary information will be delivered to the fast EMPC as well. Finally, the fast EMPC is developed based on the fast subsystem to re-optimize the fast dynamics $x_f$ in seconds. By taking into account the performance indices $J_1$ and part of $J_3$ ($u_m$ in $J_3$ is omitted), the fast dynamics feedback $x_f$ and the optimal values from the medium EMPC, the optimal fast inputs $u_f$ can be obtained and applied to the operating units that depend on the fast dynamics $x_f$.

In addition, taking into account the aforementioned thermal comfort, the building temperature set-point is allowed to change within a small range, which forms a zone tracking optimization problem. The zone tracking of building temperature not only contributes significantly to reducing the operation cost, but is also capable of giving the system more control degrees of freedom to prioritize satisfying the electric demand. In the IES, since the building temperature depends on the slow dynamics, the thermal comfort is integrated into the long-term EMPC and slow EMPC by a zone tracking constraint and a zone tracking cost, respectively. The detailed procedure of the CEMPC design will be presented in the following subsections.

\subsection{Proposed long-term EMPC with zone tracking}
This subsection presents the design of the long-term EMPC with zone tracking based on the simplified slow subsystem model. Considering the desired sampling time and prediction horizon in the long-term EMPC and the optimization problem to be solved, the simplified slow subsystem model is first developed via the reduced-order slow subsystem in Section 3.2. Next, the long-term EMPC with zone tracking is designed using the simplified model to achieve optimal start-ups/shut-downs of the operating units and the optimal capacity states of the energy storage units while taking into account the thermal comfort.

\subsubsection{Simplified slow subsystem model for the long-term EMPC}
Generally, the long-term optimization in IESs will optimize the system operation during a very long time horizon, such as 24 hours or even longer. Only approximate forecasts for future environmental conditions and customers' demands can be used in this time scale. In this situation, the long-term optimization mainly aims to determine the approximate operational states of the operating units, improve the system's economic performance and ensure the started operating units can satisfy the customers' demands in a long-term view. A simplified model that can reflect the energy balance and slow dynamics in the IES is sufficient to deal with this problem and avoid unnecessary computational costs.

In the proposed long-term EMPC, the sampling time and prediction horizon are set as 1 hour and 24 hours, respectively. The medium and fast dynamics in the operating units can be omitted in this time scale. Therefore, the slow subsystem model Eq.\eqref{66} is suitable for the long-term EMPC. Furthermore, since the long-term EMPC is a mixed-integer optimization problem, the change of the chilled water flow rates in the absorption and electric chiller, $G_{ab}$ and $G_{ec}$, and the change of the supplied/return water temperature are ignored to reduce the computational complexity. The states $C_{stc}$ and $C_{sth}$ can be omitted from the slow dynamics correspondingly due to these states depending on the chilled water temperature. Based on the above assumptions, the slow subsystem model is further linearized and rewritten for the purpose of simplicity as follows:
\begin{equation}
	\frac{d C_{soc}}{d\tau} = k_{ba}P_{ba} \label{46}
\end{equation}
\begin{equation}
	\frac{d C_{sot}}{d\tau} = -\frac{G_{st}}{M_{st}}
\end{equation}
\begin{equation}
	\frac{d t_{br}}{d\tau} = \frac{\alpha_{br}A_{br}(t_a - t_{br}) - Q_{sl} + Q_{other}}{C_{mbr}}
\end{equation}
where $k_{ba}$ is a conversion coefficient; Eq.\eqref{46}, the capacity state of the battery $C_{soc}$, is obtained by linearizing Eq.\eqref{37} at $P_{ba} = 0$. The capacity state of the cold tank $C_{sot}$ and the building temperature $t_{br}$ keep the original expressions. Note that the dynamics of the building $t_{br}$ remains in the simplified slow subsystem model. For the other operating units, the linearized power output expressions are presented below:
\begin{equation}
	\begin{aligned}
		&P_{pv}=k_{pv}S_{ra},\ \ P_{fc}=k_{fc}G_{ff},\ \ 	P_{mt}=k_{mt}G_{fm},\ \ Q_{ab}=k_{ab}P_{mt} \\
		&Q_{ec}=k_{ec}N_{ec},\ \ P_{cp}=k_{cp}Q_{ec},\ \ Q_{st}=k_{st}G_{st},\ \ P_{pmp} = k_{pmp}G_{all} \label{49}
	\end{aligned}
\end{equation}
where the subscripts $pv$, $fc$, $mt$, $ab$, $ec$, $cp$, $st$, $pmp$ represent the photovoltaic energy generation unit, fuel cell, microturbine, absorption chiller, electric chiller, compressor, chilled water storage unit, and water pump, respectively; $k$ in Eq.\eqref{49} represents conversion factors in each operating unit and are evaluated under their nominal conditions; and $G_{st}$ is still the water flow rate of the tanks which is computed by $G_{st} = z_{st}G_{stu} + (z_{st} - 1)G_{stu}$; $G_{all}$ is the total water flow rate and evaluated by $G_{all} = z_{ma}G_{ab,0} + z_{ec}G_{ec,0} + G_{stu}$, where $G_{ab,0}$ and $G_{ec,0}$ are the constants that represent the chilled water flow rate in the absorption and electric chiller under the nominal condition. Then, the electric and cooling power supplied by the IES can be rewritten as follows:
\begin{equation}
	P_{sl}=P_{pv} + P_{fc} + P_{mt} + P_{ba} - P_{cp} - P_{pmp}
\end{equation}
\begin{equation}
	Q_{sl}=Q_{ab} + Q_{ec} + Q_{st} \label{58}
\end{equation}

The above equations Eqs.\eqref{46}-\eqref{58} constitute the simplified slow subsystem model of the IES for long-term EMPC. According to the simplified model, define the state vector for long-term optimization as $x_l = [C_{soc}, C_{sot}, t_{br}]^T$, the continuous manipulated inputs for long-term as $u_l = [G_{ff}, G_{fm}, N_{ec}, G_{stu}, P_{ba}]^T$, the 0-1 integer manipulated inputs as $	z_l = [z_{fc}, z_{ma}, z_{ec}, z_{st}]^T$, the disturbances vector for long-term as $\omega_l = [t_a, S_{ra}, Q_{other}]^T$, the output vector for long-term as $y_l = [P_{sl}, t_{br}]^T$. Then the simplified slow subsystem model of the IES is rewritten as a compact form as follows:
\begin{equation}
	\begin{aligned}
		&\dot{x_l}(\tau) = f_l(x_l(\tau),u_l(\tau),z_l(\tau),\omega_l(\tau))\\
		&y_l(\tau) = h_l(x_l(\tau),u_l(\tau),z_l(\tau),\omega_l(\tau)) \label{59}
	\end{aligned}
\end{equation}
where $x_l\in \mathbb{R}^{3}$, $u_l\in \mathbb{R}^{5}$, $z_l\in \{0,1\}^4$, $\omega_l \in \mathbb{R}^{3}$ and $y_l\in \mathbb{R}^{2}$.
\subsubsection{Design of long-term EMPC}
In order to optimize the start-ups/shut-downs of the operating units and the operational trajectories of energy storage units, ensure the energy balance, reduce the operating cost and achieve the thermal comfort, a long-term EMPC with zone tracking based on the simplified slow subsystem is proposed for the IES as follows:
\begin{subequations} \label{60}
	\begin{align}
		\min_{u_l,x_l,z_l,y^l_{\epsilon}} & \sum_{i=1}^{N^l_p}
		\alpha^l(u_{l,1}(k+i-1)+u_{l,2}(k+i-1)) + \|y^l_\epsilon(k+i)\|^2_{R^l} \label{60a}\\
		s.t.\ &x_l(k+i) = f_l(x_l(k+i-1),u_l(k+i-1),z_l(k+i-1),\omega_l(k+i-1)) \label{60b}\\
		&y_l(k+i)=h_l(x_l(k+i),u_l(k+i),z_l(k+i),\omega_l(k+i)) \label{60c}\\
		&z^l_u(k+i-1) u^l_{min}\leq u_l(k+i-1)\leq z^l_u(k+i-1) u^l_{max} \label{60d}\\
		&du^l_{min}\cdot\Delta_l\leq u_l(k+i-1) - u_l(k+i-2)\leq du^l_{max}\cdot\Delta_l \label{60h}\\
		&x^l_{min}\leq x_l(k+i)\leq x^l_{max} \label{60e}\\
		&0 = y_{l,1}(k+i) - y^l_{sp,1}(k+i) + y^l_{\epsilon,1}(k+i) \label{60f}\\
		&y^l_L(k+i) - y^l_{\epsilon,2}(k+i)\leq y_{l,2}(k+i) \leq y^l_H(k+i) + y^l_{\epsilon,3}(k+i) \label{60g}
	\end{align}
\end{subequations}
where $\alpha^l$ and $R^l$ are weighting coefficient and weighting matrix; $ y^l_{sp,1}(k)$ is the set-point for the first output in Eq.\eqref{59}, i.e., the electric demand; $z^l_u(k)$ is a matrix composed of the integer variables and defined as $z^l_u(k) = {\rm diag}([z_{l,1}(k), z_{l,2}(k), z_{l,3}(k), 1, 1])$; $N^l_p$ represents the prediction horizon, and $i=1,\dots, N^l_p$; $\Delta_l$ is the long-term sampling time; $y^l_\epsilon(k)$ is a vector of the slack variables introduced to improve the feasibility of optimization problems; $y^l_L(k)$ and $y^l_H(k)$ are presented to realize zone tracking control and used to define the lower and upper bound of the target zone for the controlled output, which is expressed as:
\begin{subequations}
	\begin{align}
		&y^l_L(k+i) = y^l_{sp,2}(k+i) - \delta(k+i) \\
		&y^l_H(k+i) = y^l_{sp,2}(k+i) + \delta(k+i)
	\end{align}
\end{subequations}
where $y^l_{sp,2}(k)$ represents the original set-point for the second output in Eq.\eqref{59} i.e., building temperature, and $\delta(k)$ is the predetermined relaxation value from the original set-point. By setting $\delta(k)$ to a non-zero value, the proposed long-term EMPC will achieve the goal of the zone tracking which will allow for variations in the outputs within the target zone and make the entire system more flexible and economic.

In the long-term EMPC Eq.\eqref{60}, Eq.\eqref{60a} defines the objective function that minimizes the fuel consumption of daily operation and output target tracking; Eqs.\eqref{60b}-\eqref{60c} are the model constraints, which is attained by discretizing Eq.\eqref{59} with discrete step $\Delta_l$ that equals sampling time; Eqs.\eqref{60d}-\eqref{60e} are the manipulated variable and state constraints; because the EMPC optimizes the output trajectories in a discrete time fashion, the set-point tracking of the electric demand and the zone tracking of the building temperature are expressed in the form of constraints Eq.\eqref{60f} and Eq.\eqref{60g} for maintaining the energy balance between the system and customers within each long-term sampling interval $\Delta_l$. At each sampling time $k$, all of the optimal decision variable sequences, $\{u_l^*(k),\dots, u_l^*(k+N^l_p-1)\}$, $\{z_l^*(k),\dots, z_l^*(k+N^l_p-1)\}$, $\{x_l^*(k+i),\dots, x_l^*(k+N^l_p)\}$ and $\{y_\epsilon^*(k+i),\dots, y_\epsilon^*(k+N^l_p)\}$, are calculated simultaneously by solving this optimization problem. Then the first integer manipulated input, $z_l^*(k)$, is applied to the system and the first optimal operational trajectories of the energy storage units, $C_{soc}^*(k+1) = x_{l,1}^*(k+1)$ and $C_{sot}^*(k+1) = x_{l,2}^*(k+1)$, are delivered to the short-term EMPC as a reference for regulating the battery and chilled water storage unit. At next sampling time $k+1$, the long-term EMPC is reinitialized and computes new optimal sequences. Here to facilitate the introduction of short-term EMPC later, let us define this $z^l = z_l^*(k)$, $C_{soc}^l = C_{soc}^*(k+1)$ and $C_{sot}^l = C_{sot}^*(k+1)$. It should be noted that the optimized continuous inputs $u^*_l$ in the long-term EMPC are discarded. The complete continuous inputs will be re-optimized in the short-term EMPC by exploiting actual external conditions and precise system models.

\subsection{Proposed short-term EMPC with zone tracking}
From the perspective of the short-term operation optimization of the IES, the primary goal is to coordinate the system in real-time to satisfy the customers' electric demand while maintaining the building temperature within an acceptable range and advancing the energy efficiency. To this end, a short-term EMPC with zone tracking based on multi-time-scale decomposition is proposed in this subsection while overcoming the problem of large-scale nonlinear optimization with time-scale multiplicity. In the proposed design, the short-term EMPC consists of three interconnected EMPCs founded on different subsystems, i.e., slow EMPC, medium EMPC, and fast EMPC, to capture the multi-time-scale dynamics featured in the IES. A zone tracking cost is incorporated into the slow EMPC to achieve the thermal comfort. The optimal operational trajectories of the energy storage units and the decisions about the start-ups/shut-downs of the operating units from the long-term EMPC are taken into consideration in the short-term EMPC as well.
\subsubsection{Design of slow EMPC}
Taking advantage of system decomposition and zone tracking, the proposed slow EMPC optimization for the reduced-order slow subsystem is expressed as follows:
\begin{subequations} \label{72}
	\begin{align}
		\min_{u_s, u_m, u_f, y_{sp,\epsilon}} &\sum_{i=1}^{N^s_p} \alpha^s_1\|y_{mf}(k+i) - y_{sp,1}(k+i)\|^2 + \alpha^s_2\|y_s(k+i) - y_{sp,\epsilon}(k+i)\|^2 \notag\\&+ \alpha^s_3(u_m(k+i-1)+u_{f,1}(k+i-1)) \notag\\ &+ \alpha^s_4\|x_{s,1}(k+i) - C_{soc}^l(k+i)\|^2 + \alpha^s_5\|x_{s,2}(k+i) - C_{sot}^l(k+i)^l\|^2 \label{72a}\\
		s.t.\ &x_s(k+i) = f_{sc}(x_s(k+i-1),u_s(k+i-1),u_m(k+i-1),\notag\\ & u_f(k+i-1),z^l(k+i-1),\omega(k+i-1)) \label{72c}\\ &y_s(k+i)=h_s(x_s(k+i)) \label{72d}\\ &y_{mf}(k+i)=h^s_{mf}(x_s(k+i),u_s(k+i),u_m(k+i),u_f(k+i), \notag \\ &z^l(k+i),\omega(k+i)) \label{72e}\\ &z^j_u(k+i-1) u^j_{min}\leq u_j(k+i-1)\leq z^j_u(k+i-1) u^j_{max} \label{72f}\\ &du^j_{min}\cdot\Delta_s\leq u_j(k+i-1) - u_j(k+i-2)\leq du^j_{max}\cdot\Delta_s \label{72g}\\
		&x^j_{min}\leq x_j(k+i)\leq x^j_{max} \label{72h}\\ &y^s_L(k+i)\leq y_{sp,\epsilon}(k+i) \leq y^s_H(k+i) \label{72i}
	\end{align}
\end{subequations}
where $N^s_p$ is the prediction horizon and $i=1,\dots, N^s_p$; $\alpha^s$ is the weighting vector; $y_{sp,1}(k)$ is the set-point of the first output, i.e., the demand for electric power, in Eq.\eqref{42}; $y_{sp,\epsilon}(k)$ is the zone variable introduced to realize the zone tracking control for building temperature; $\Delta_s$ is the slow sampling time; $j$ is a index to indicate a variable correlated with the the slow, medium and fast dynamics and $j = s,m,f$; $z^j_u$ is used to describe matrices made up of $z^l(k)$, specifically, $z^s_u(k) = {\rm diag}([z^l_2(k),z^l_3(k),1])$, $z^m_u(k) = z^l_2(k)$, $z^f_u(k) = {\rm diag}([z^l_1(k),z^l_3(k)])$; $C_{soc}^l(k)$, $C_{sot}^l(k)$ and $z^l(k)$ are obtained from the long-term EMPC; $y^s_L(k)$ and $y^s_H(k)$ represent the lower and upper bound of target zone for the controlled variable for the building temperature and the expression is as follows:
\begin{subequations}
	\begin{align}
		&y^s_L(k+i) = y_{sp,2}(k+i) - \delta(k+i) \\
		&y^s_H(k+i) = y_{sp,2}(k+i) + \delta(k+i)
	\end{align}
\end{subequations}
where $y_{sp,2}(k)$ indicates the original set-point for building temperature and $\delta(k)$ serves as the same predetermined relaxation value from the original set-point as long-term EMPC. The proposed slow MPC will also achieve the zone tracking control by setting $\delta(k)$ to a non-zero value which will overlook the output variations within the target zone and make the entire system more robust and less sensitive to environmental uncertainties. On the contrary, by setting $\delta(k)$ to zero the controller will achieve a precisely tracking for set-point.

In this optimization problem, Eq.\eqref{72a} defines the optimization objective function that needs to be minimized. In the objective function Eq.\eqref{72a}, the first and second terms is used to realize the set-point tracking for the electric demand and the zone tracking for the building temperature, respectively; the third term takes into account the economic performance of the IES, saving fuel as far as possible; the last two terms make sure that the battery bank and chilled water storage unit operation is in accordance with the optimal trajectories from the long-term EMPC. Eqs.\eqref{72c}-\eqref{72e} describe the model constraints on the reduced-order slow subsystem, which is obtained from discretization for Eq.\eqref{67} with discrete step $\Delta_s$ that equals sampling time. Eqs.\eqref{72f}-\eqref{72h} are the manipulated variable and state constraints respectively. Where the constraint on the inputs changing rates Eq.\eqref{72g} is valid during equipment operating normally and will be ignored at the finite time instant when equipment is switched on/off. Eq.\eqref{72i} represents the zone tracking constraints on the zone variable. At each sampling time $k$, all of the optimal decision variable sequences, $\{u_s^*(k),\dots, u_s^*(k+N^s_p-1)\}$, $\{u_m^*(k),\dots, u_m^*(k+N^s_p-1)\}$, $\{u_f^*(k),\dots, u_f^*(k+N^s_p-1)\}$ and $\{y_{sp,\epsilon}^*(k+i),\dots, y_{sp,\epsilon}^*(k+N^s_p)\}$, are calculated simultaneously by solving this optimization problem. Then the first optimal slow manipulated input, $u_s^*(k)$, applies to the IES. The first optimal medium and fast manipulated input, $u_m^*(k)$ and $u_f^*(k)$, and the correspondingly optimal medium and fast state, $x_m^*(k+1)$ and $x_f^*(k+1)$ evaluated by Eqs.\eqref{66b} and \eqref{66c}, are sent to the medium EMPC as reference. At next sampling time $k+1$, the slow MPC will re-implement and compute new optimal sequences. In order to facilitate the subsequent medium and fast EMPC's design, let us define $u^s_s=u_s^*(k)$, $u^s_m=u_m^*(k)$, $u^s_f=u_f^*(k)$, $x^s_m=x_m^*(k+1)$ and $x^s_f=x_f^*(k+1)$.

\subsubsection{Design of medium EMPC}
From the reduced-order medium subsystem, the proposed medium EMPC optimization is formulated as follows:
\begin{subequations} \label{74}
	\begin{align}
		\min_{u_m, u_f}\ &\sum_{i=1}^{N^m_p} \alpha^m_1\|y_{mf}(k+i) - y_{sp,1}(k+i)\|^2 + \alpha^m_2(u_m(k+i-1)+u_{f,1}(k+i-1)) \notag\\ &+ \|x_m(k+i) - x^s_m(k+i)\|^2_{R^m_1} + \|u_m(k+i-1) - u^s_m(k+i-1)\|^2_{R^m_2} \notag\\ &+ \|x_f(k+i) - x^s_f(k+i)\|^2_{R^m_3} + \|u_f(k+i-1) - u^s_f(k+i-1)\|^2_{R^m_4} \label{74a}\\
		s.t.\ &x_m(k+i) = f_{mc}(x_s(k+i-1),x_m(k+i-1),u^s_s(k+i-1),\notag\\ & u_m(k+i-1),u_f(k+i-1),z^l(k+i-1)) \label{74c}\\ &y_{mf}(k+i)=h^m_{mf}(x_m(k+i),u^s_s(k+i),u_f(k+i), z^l(k+i),\omega(k+i)) \label{74d}\\ &z^j_u(k+i-1) u^j_{min}\leq u_j(k+i-1)\leq z^j_u(k+i-1) u^j_{max} \label{74e}\\ &du^j_{min}\cdot\Delta_m\leq u_j(k+i-1) - u_j(k+i-2)\leq du^j_{max}\cdot\Delta_m \label{74f}\\
		&x^j_{min}\leq x_j(k+i)\leq x^j_{max} \label{74g}
	\end{align}
\end{subequations}
where $N^m_p$ is the prediction horizon and $i=1,\dots, N^m_p$; $\alpha^m$ and $R^m$ are the weighting vector and the weighting matrix; $y_{sp,1}(k)$ here is the same set-point as in the slow EMPC; $\Delta_m$ is the medium sampling time; $j$ is a index to indicate a variable associated with the medium and fast dynamics and here $j = m,f$; $z^m_u(k)$ $z^f_u(k)$ and $z^l(k)$ are the same integer variables as in the slow EMPC; and $u^s_s(k)$ is obtained from the slow EMPC; $x_s(k)$ will be held a constant for slow dynamics in Eq.\eqref{74} according to Eq.\eqref{68a} and updated every slow sampling time $\Delta_s$.

In the optimization problem of Eq.\eqref{74}, Eq.\eqref{74a} is the cost function in which the first two terms are employed to track the electric demand while saving fuel during medium time-scale transient; and the cost function utilizes the last four terms to ensure its results are consistent in the optimal value from the slow EMPC eventually. Eqs.\eqref{74c} and \eqref{74d} represent the model constraints on the reduced-order medium subsystem and obtained from Eq.\eqref{69} discretization with discrete step $\frac{\Delta_m}{\epsilon_2}$. Eqs.\eqref{74e}-\eqref{74g} serve as the input variable and state constraints. Similarly, at each sampling time k, all of the optimal decision variable sequences, $\{u_m^*(k),\dots, u_m^*(k+N^m_p-1)\}$ and $\{u_f^*(k),\dots, u_f^*(k+N^m_p-1)\}$, are obtained simultaneous. Subsequently, the first optimal medium manipulated input $u^*_m(k)$ enters the IES. The first fast manipulated input $u^*_f(k)$ and correspondingly optimal fast state $x^*_f(k+1)$ calculated by \eqref{68c} are forwarded to the fast EMPC as reference. At the next sampling time $k+1$, the above optimization process will be implemented again. Let us define $u^m_f(k) = u^*_f(k)$ and $x^m_f = x^*_f(k+1)$ for the fast EMPC.

\subsubsection{Design of fast EMPC}
To proceed, the proposed fast EMPC based on the reduced-order fast subsystem is described as follows:
\begin{subequations} \label{75}
	\begin{align}
		\min_{u_f}\ &\sum_{i=1}^{N^f_p} \alpha^f_1\|y_{mf}(k+i) - y_{sp,1}(k+i)\|^2 + \alpha^f_2 u_{f,1}(k+i-1) \notag\\ &+ \|x_f(k+i) - x^m_f(k+i)\|^2_{R^f_1} + \|u_f(k+i-1) - u^m_f(k+i-1)\|^2_{R^f_2} \label{75a}\\
		s.t.\ &x_f(k+i) = f_{fc}(x_m(k+i-1),x_f(k+i-1),u^s_s(k+i-1),\notag\\ &u_f(k+i-1),z^l(k+i-1),\omega(k+i-1)) \label{75c}\\ &y_{mf}(k+i)=h^f_{mf}(x_m(k+i),x_f(k+i),u^s_s(k+i),u_f(k+i),\notag \\ &z^l(k+i),\omega(k+i)) \label{75d}\\ &z^f_u(k+i-1) u^f_{min}\leq u_f(k+i-1)\leq z^f_u(k+i-1) u^f_{max} \label{75e}\\ &du^f_{min}\cdot\Delta_f\leq u_f(k+i-1) - u_f(k+i-2)\leq du^f_{max}\cdot\Delta_f \label{75f}\\
		&x^f_{min}\leq x_f(k+i)\leq x^f_{max} \label{75g}
	\end{align}
\end{subequations}
where $\Delta_f$ is the fast sampling time; the rest of mathematical symbol in Eq.\eqref{75} is similar to the slow EMPC's and medium EMPC's and will not be repeated here. Please note that, according to Eq.\eqref{70b}, $x_m(k)$ holds as constants for medium dynamics in Eq.\eqref{75} until next medium sampling time $\Delta_m$ comes; and $u^s_s$ is sent from the slow EMPC.

In the fast EMPC's optimization problem, Eq.\eqref{75a} is the expressions of the objective function, which realizes the set-point tracking for the electric demand, and improving economic performance during the fast time-scale dynamics by minimizing the first two terms. And the last two terms are used to guarantee the optimization results in keeping with the medium EMPC eventually. Eqs.\eqref{75c} and \eqref{75d} express the model constraints on the reduced-order fast system and achieved from discretization for Eqs.\eqref{71} with discrete step $\frac{\Delta_f}{\epsilon_1}$. Eqs.\eqref{75e}-\eqref{75g} are presented for the input and state constraints. At sampling time $k$, the optimal sequence $\{u_f^*(k),\dots, u_f^*(k+N^f_p-1)\}$ is obtained through solving Eq.\eqref{75} and the first optimal fast manipulated input $u_f^*(k)$ is implemented in the IES, which optimization process will be repeated when $k+1$ comes.
\begin{remark}
	Note that the short-term EMPC for the IES possesses the different sampling time that in the slow EMPC of Eq.\eqref{72}, the control action is calculated every $\Delta_s$; in the medium EMPC of Eq.\eqref{74}, the control action is calculated every $\Delta_m$; in the fast EMPC of \eqref{75}, the control action is calculated every $\Delta_f$; and in whole short-term EMPC Eqs.\eqref{72}, \eqref{74} and \eqref{75}, the slow state $x_s$ is sampled every $\Delta_s$, the medium state $x_m$ every $\Delta_m$, the fast state $x_f$ every $\Delta_f$.
\end{remark}

\begin{remark}
	It should be noted that an individual subsystem contains different operating units and an individual operating unit may have different time scales. Consequently, an individual EMPC in the short-term EMPC may regulate one or more operating units, and an operating unit may also be controlled by different EMPCs in which the communication ensures the consistency of optimization results as shown in Figure \ref{add_f1}. The design of this control scheme depends on which kind of time-scale dynamics the operating unit has and how the operating units are interconnected. Besides, one manipulated input that is simultaneously associated with different time scales will be optimized by the slower EMPC for the stability of the IES.
\end{remark}

\section{Simulation results}
In this section, we apply the proposed composite EMPC to the stand-alone IES and compare its performance with hierarchical real-time optimization approach to demonstrate the effectiveness of the proposed method. The optimization problem is solved in Python based on CasADi using the BONMIN and IPOPT solvers \cite{r50} on a machine with 16GB RAM and 2.60GHz Intel Core i7-10750H.

\subsection{Simulation parameters and hierarchical real-time optimization}

For the stand-alone IES described in Eq.\eqref{42}, the capacities of different operating units are displayed in Table \ref{t1}, and the main model parameters used in the simulation have been given in Table \ref{t2} in Section 2.1.

\begin{table}[t] \small
	\centering
	\caption{Capacities of operating units in the IES}
	\label{t1}	
	\renewcommand{\arraystretch}{1.2}
	\tabcolsep 4pt
	\begin{tabular}{cccccc} \hline
		\multicolumn{2}{c}{Electricity supply} &
		\multicolumn{2}{c}{Cooling supply} & 
		\multicolumn{2}{c}{Electricity consumption in cooling}\\
		Operating unit & Power & Operating unit & Power & Operating unit & Power \\ \hline
		Photovoltaic energy & 45 kW & Electric chiller & 50 kW & Compressor in chiller & 12.6 kW\\
		Microturbine & 80 kW & Absorption chiller & 75 kW & Chilled water pump & 13.9 kW\\
		Fuel cell & 40 kW & Chilled water storage & 117kWh & &\\
		Battery bank & 156 kWh & & & &\\ \hline
	\end{tabular}
\end{table}

For original model in Eq.\eqref{42}, the lower and upper bounds of the manipulated inputs are $u_{min} = [0.00055,0.002,1.9,20,1.3,0]^T$ and $u_{max} = [0.0045,0.0066911,3.5714,110,2.381,1]^T$, respectively. The lower and upper bounds of the changing rates of the manipulated inputs are $du_{min} = [-0.00395,-0.001173,-0.418,-70,-0.271,-0.25]^T$ and $du_{max} = [0.00395,0.001173, \\ 0.418,70,0.271,0.25]^T$. The lower and upper bounds of system states are $x_{min} = [13,0.03,0.006, \\ 0.095,0.002,-75,-0.5,-4,-5,27,25,25,-5,0,7,-0.19,0.1,-145,0.1,2000,0,8,18]^T$ and $x_{max} = [130,0.4,0.06,1,0.08,5.00,3,1,10,44,38,35,20,22,22,0.19,0.9,155,1,240000,324000,18,26]^T$, respectively. The lower and upper bounds of the controlled outputs are $y_{min} = [0,18]^T$ and $y_{max} = [195,26]^T$, respectively. The integer variable $z$ satisfies $z\in \{0,1\}^4$. The simplified long-term model's bounds can also be derived from the above original bounds, which will not be repeated.

\begin{table}[t] \small
	\centering
	\caption{Sampling time and prediction horizon in the EMPCs}
	\label{t3}
	\renewcommand{\arraystretch}{1.2}
	\tabcolsep 10pt
	\begin{tabular}{ccc} \hline
		EMPCs& Sampling time & Prediction horizon \\ \hline
		Long-term EMPC & $\Delta_l = 1\ hr$& $N^l_p = 24$ \\
		Slow EMPC & $\Delta_s = 120\ s$& $N^s_p = 12$ \\
		Medium EMPC & $\Delta_m = 8\ s$& $N^m_p = 12$ \\
		Fast EMPC & $\Delta_f = 2\ s$& $N^f_p = 4$ \\ \hline
	\end{tabular}
\end{table}

In the remainder of this work, considering external conditions and system dynamics, the sampling time and prediction horizon in EMPCs are given in Table \ref{t3}, and the weighting parameters in each EMPC are chosen as follows: for long-term EMPC, the weighting parameter and matrix are chosen as $\alpha^l = 1$ and $R^l = I_{3\times3}$ where $I$ serves as identity matrix; for slow EMPC in short-term, the weighting parameters are chosen as $\alpha^s_1 = 5$, $\alpha^s_2= 5$, $\alpha^s_3 = 5$, $\alpha^s_4 = 8$, $\alpha^s_5 = 8$; for medium EMPC, the weighting parameters and matrices are chosen as $\alpha^m_1 = 8$, $\alpha^m_2= 1$, $R^m_1 = 8\cdot I_{9\times9}$, $R^m_2 = 8\cdot I_{1\times1}$, $R^m_3 = 8\cdot I_{9\times9}$, $R^m_4 = 8\cdot I_{2\times2}$; for fast EMPC, the weighting parameters and matrices are chosen as $\alpha^f_1 = 8$, $\alpha^f_2= 1$, $R^f_1 = 8\cdot I_{9\times9}$, $R^f_2 = 8\cdot I_{2\times2}$. Note that all of the weighting parameters and matrices are calculated after normalization.

\begin{figure}[t]
	\centering
	\includegraphics[width=0.32\hsize]{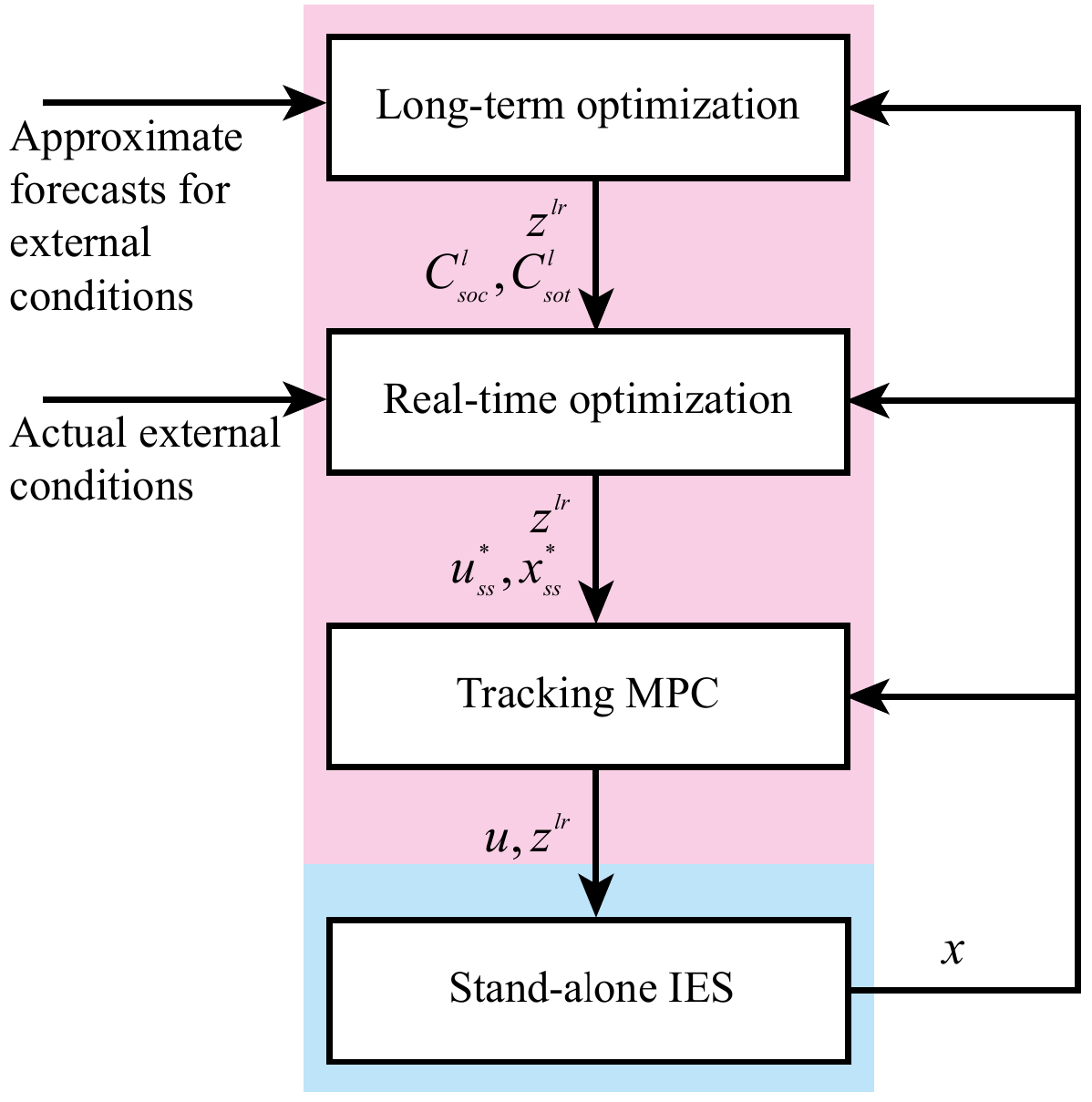}
	\caption{A hierarchical real-time optimization for comparison}
	\label{f3}
\end{figure}

Since the hierarchical optimization mechanism is capable of addressing complex optimization problems, it is an attractive solution for the optimal operation of distributed energy systems and has been extensively investigated in existing works \cite{r11,r51,r53,r13,r18,r20}. Therefore, a hierarchical real-time optimization (HRTO) is introduced for comparison purposes in this work. The HRTO consists of three layers: the upper layer of long-term optimization, the middle layer of real-time optimization, and the tracking control layer on the bottom, as shown in Figure \ref{f3}. The optimal trajectories of the energy storage units and the start-ups/shut-downs of the operating units are determined in a long-term optimization based on Eq.\eqref{59} like the proposed techniques. The optimal steady-state operating points of the entire IES are optimized in the real-time optimization layer based on Eq.\eqref{66} and then sent to the tracking control layer to apply the manipulated inputs to the system. Moreover, both long-term and real-time optimization layers incorporate receding horizon implement in keeping with the state-of-the-art \cite{r3,r51,r53,r13,r54,r21,r52}. A centralized set-point tracking MPC based on the originally dynamic model Eq.\eqref{42} is used in the tracking layer. Specifically, the objective functions of long-term $J_{lr}$ and real-time optimization $J_r$ are similar to the proposed long-term EMPC and slow EMPC as follows:
\begin{equation}
	\begin{aligned}
		J_{lr} = & \sum_{i=1}^{N^{lr}_p}
		\alpha^{lr}(u_{lr,1}(k+i-1)+u_{lr,2}(k+i-1)) + \|y_\epsilon(k+i)\|^2_{R^{lr}} \label{76}
	\end{aligned}
\end{equation}

\begin{equation}
	\begin{aligned}
		J_r = &\sum_{i=1}^{N^r_p} \alpha^r_1\|y_{ss,1}(k+i) - y_{sp,1}(k+i)\|^2 + \alpha^r_2\|y_{ss,2}(k+i) - y_{sp,2}(k+i)\|^2 \\&+ \alpha^r_3(u_{ss,1}(k+i-1)+u_{ss,2}(k+i-1)) \\ &+ \alpha^r_4\|x_{ss,17}(k+i) - C_{soc}^{lr}(k+i)\|^2 + \alpha^r_5\|x_{ss,19}(k+i) - C_{sot}^{lr}(k+i)\|^2
	\end{aligned}
\end{equation}
where subscripts and superscripts $lr$ represents the long-term optimization layer, superscripts $r$ indicates the real-time optimization layer, subscript $ss$ represents the steady-state. Above two objective functions are similar to the proposed approach and will not be discussed again. The optimization objective of the tracking control layer $J_t$ is expressed as:
\begin{equation}
	\begin{aligned}
		J_t = &\sum_{i=1}^{N^t_p} \|x(k+i) - x^*_{ss}(k+i)\|^2_{R^t_1} + \|u(k+i-1) - u^*_{ss}(k+i-1)\|^2_{R^t_2} \label{77}
	\end{aligned}
\end{equation}
where superscript $t$ indicates the tracking control layer and $*$ serves as the optimal value from the real-time optimization layer; two terms in Eq.\eqref{77} are the penalties for the states and inputs tracking error, respectively. The main difference from the proposed approach is that, first, in the HRTO without consideration of the thermal comfort, the building temperature set-point tracking is used instead of zone tracking. Second, the HRTO separates the overall coordinated control problem into the optimization problems and tracking problems. The real-time optimization layer is used to simply solve the optimal system operating points instead of directly regulating the manipulated inputs. Last, the HRTO does not take into account the dynamic time-scale multiplicity. The tracking layer completely manipulates all continuous inputs by tracking the optimal points sent from the upper layer, which cannot re-optimize the optimal points in different time-scale transient.

For the long-term and real-time optimization layer, the prediction horizon, sampling time and weight parameters are chosen as the same value as the proposed long-term EMPC and slow EMPC. For the tracking layer, the sampling time and prediction horizon are $\Delta_t = 5 s$ and $N^t_p = 12$, the weighting matrices are chosen as $R^t_1 = 8\cdot I_{23\times23}$, $R^t_2 = 8\cdot I_{6\times6}$.

\subsection{Results of IES under normal external conditions}

\begin{figure}[t]
	\centering
	\includegraphics[width=0.45\hsize]{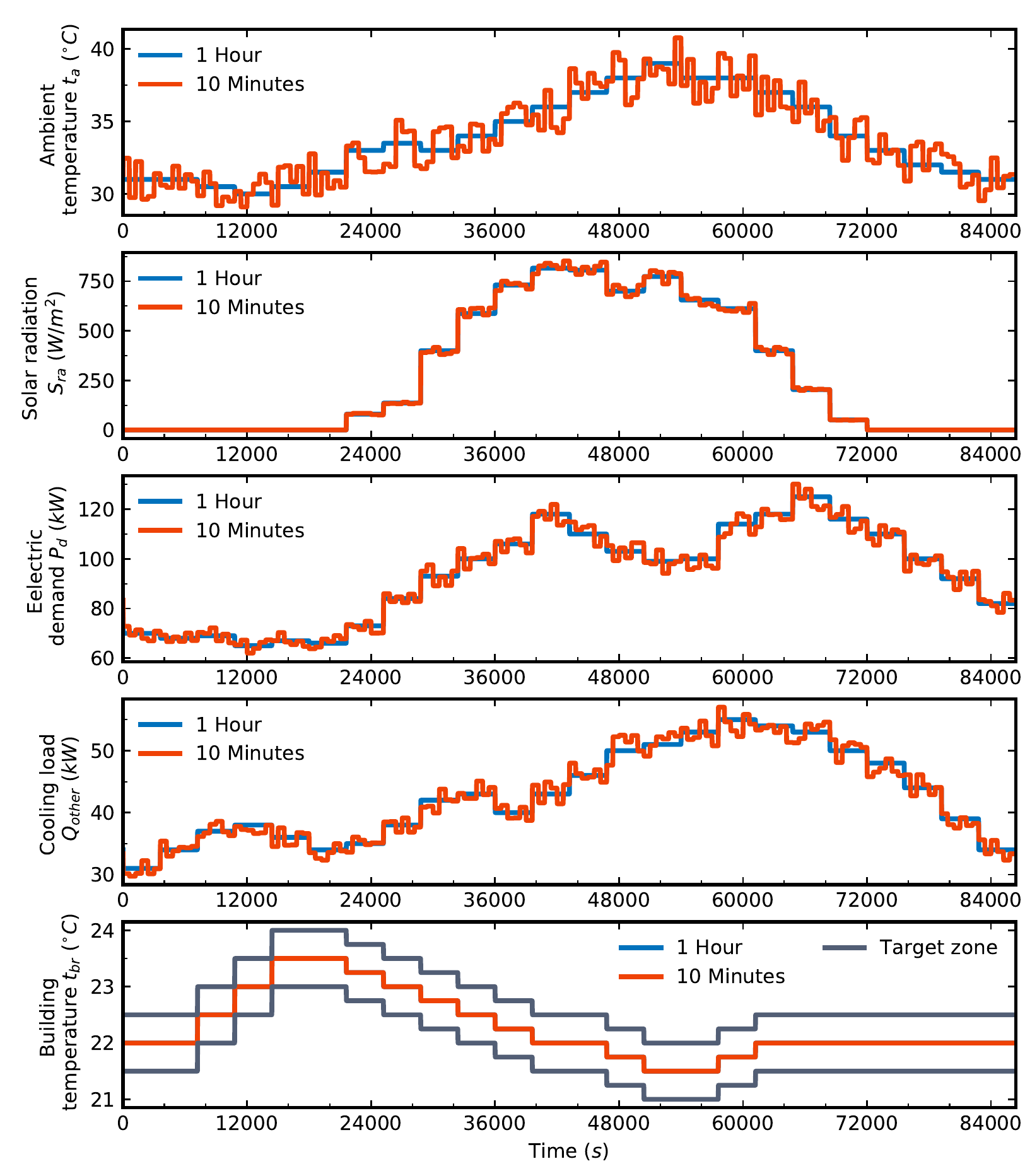}
	\caption{Normal external conditions (Scenario 1)}
	\label{f4}
\end{figure}
In this subsection, a 24-hour simulation is carried out under typically normal environmental conditions and customers' demands (Scenario 1). Figure \ref{f4} shows the continuous evolution of ambient temperature, solar radiation, customers' electric demand, other cooling loads, and building temperature demand over 24 hours. In Figure \ref{f4}, the blue lines denote the long-term environmental and demand forecasts used in the long-term optimization and are updated hourly; the red lines represent the actual environmental conditions and demand curves. The real-time curves randomly oscillate within $\pm 5\%$ around the long-term forecasts and are updated every 10 minutes except for the building temperature. As aforementioned, the control objectives are to meet the customers' electric demand firstly and maintain the building temperature within an acceptable range while saving fuel. Considering the thermal comfort, we allow the building temperature variations within $\pm 0.5^{\circ}C$ of the customers-specified temperature in the remainder of simulations. Correspondingly, the relaxation value $\delta$ is set to 0.5 in the proposed CEMPC, while the HRTO will track the customer-specified temperature.

\subsubsection{Resulting performance}

\begin{figure}[t]
	\centering
	\includegraphics[width=0.95\hsize]{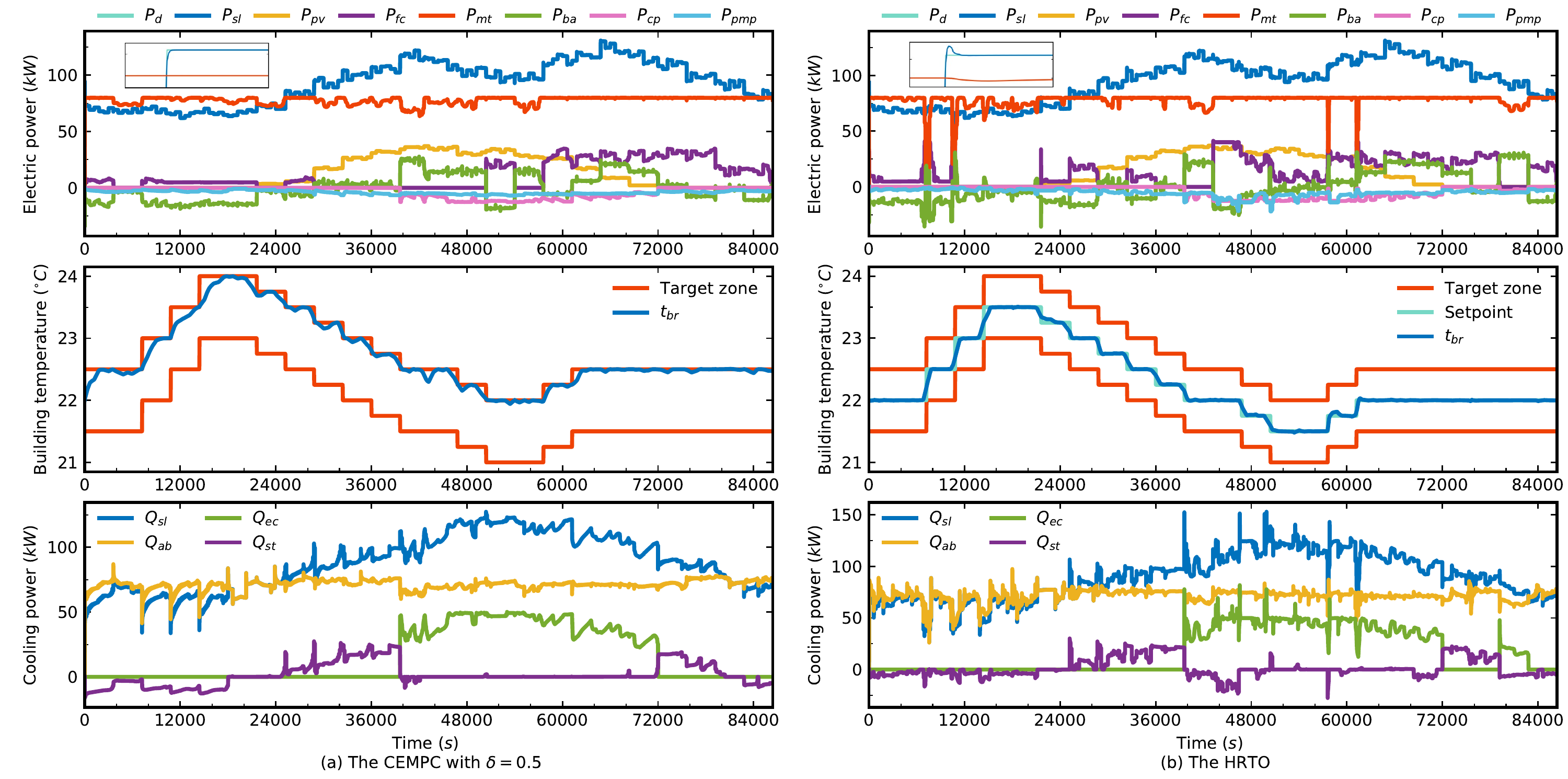}
	\caption{The outputs of the IES and power generation in each operating unit under (a) CEMPC, (b) HRTO in Scenario 1}
	\label{f5}
\end{figure}
\begin{figure}[t]
	\centering
	\includegraphics[width=0.6\hsize]{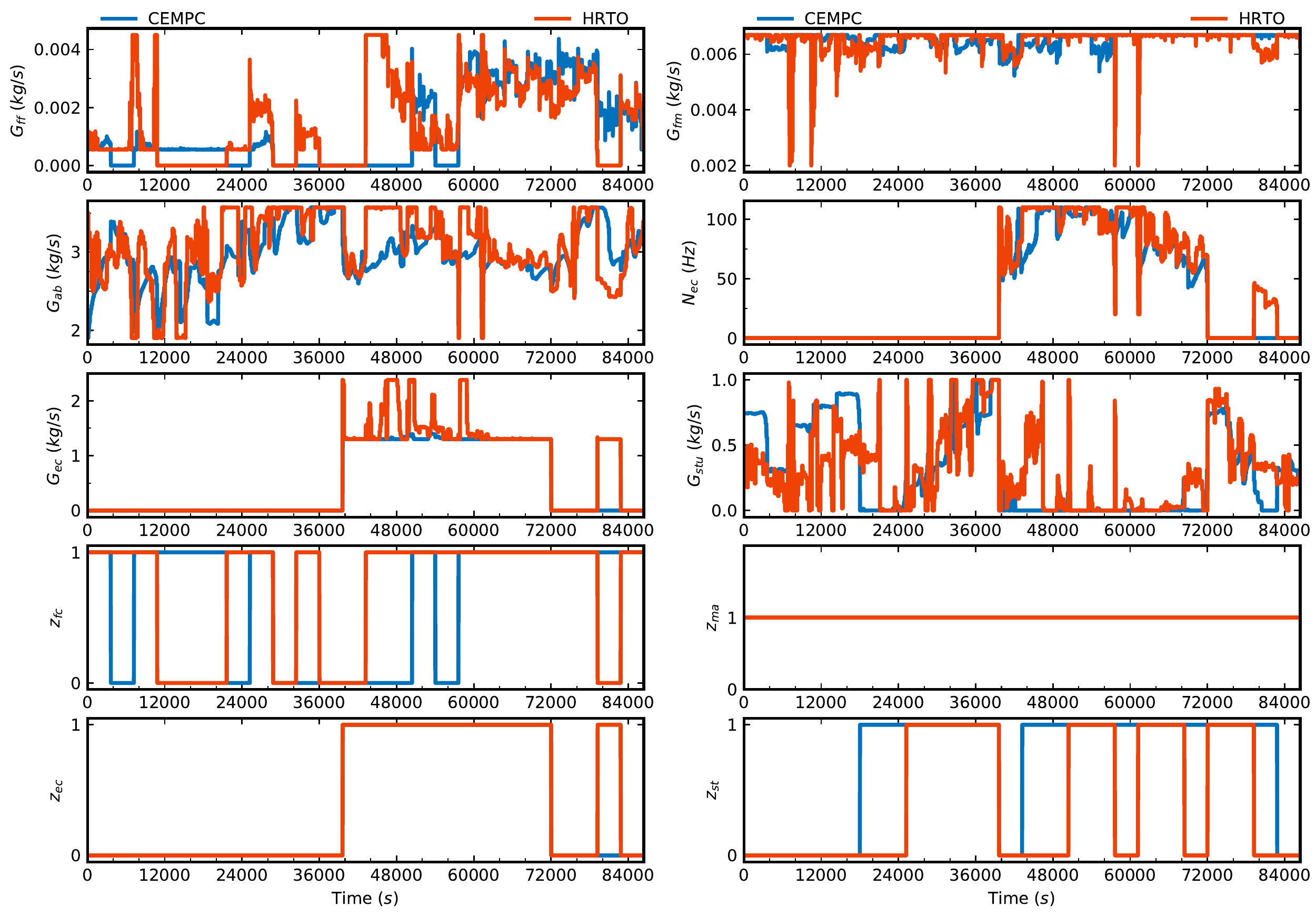}
	\caption{The manipulated inputs of the IES in Scenario 1}
	\label{f7}
\end{figure}
\begin{figure}[h]
	\centering
	\includegraphics[width=0.4\hsize]{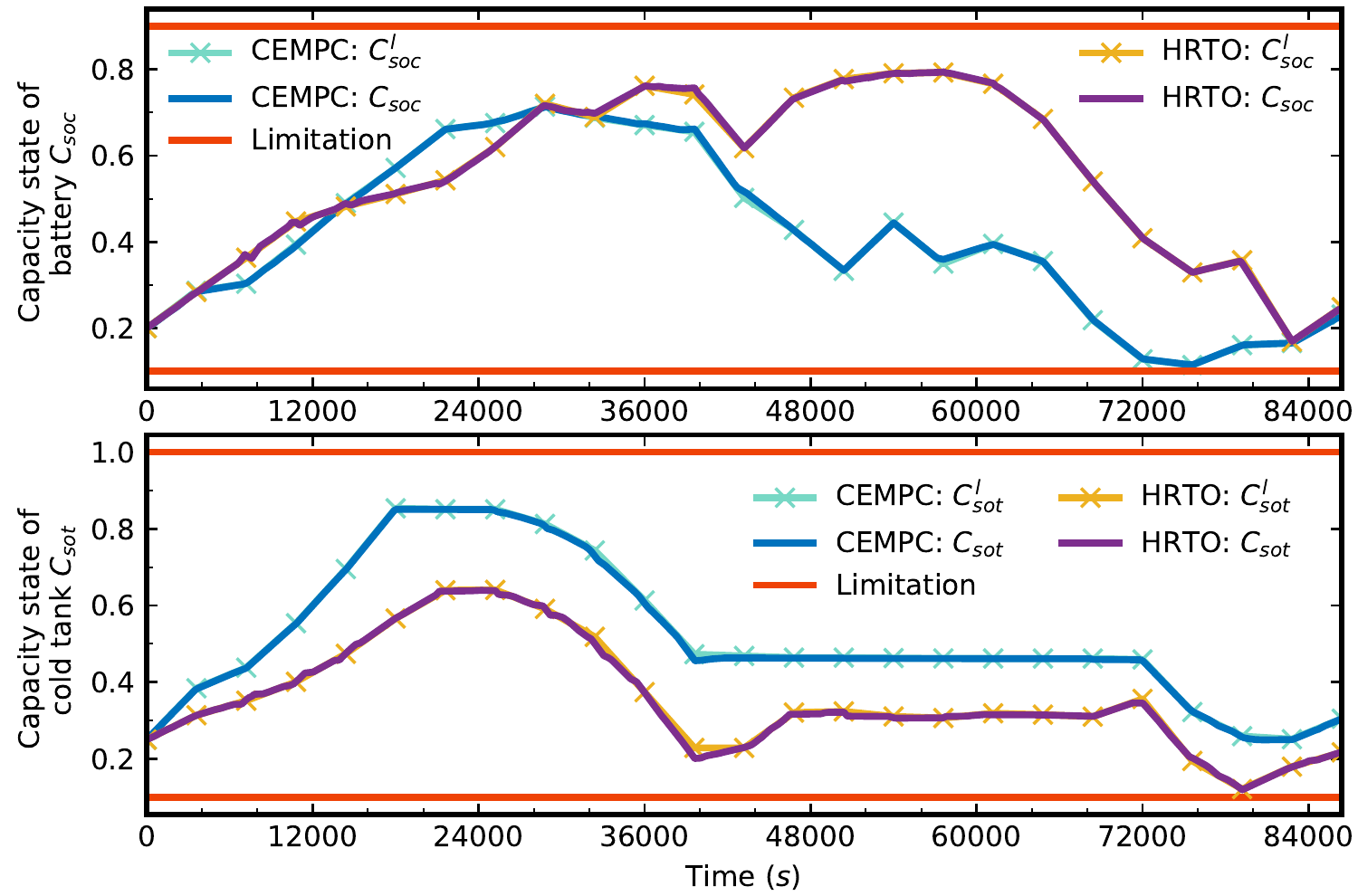}
	\caption{The capacity states of the battery bank and cold tank in Scenario 1}
	\label{f8}
\end{figure}
Figure \ref{f5} depicts the resulting output variables of the IES and the electric and/or cooling power in each operating unit under the CEMPC and HRTO, and Figure \ref{f7} displays the shape of the corresponding input variables. By comparing the results shown in Figure \ref{f5}, it can be seen that on the one hand, both the proposed CEMPC and HRTO can track the electric demand under the various environment and demands. However, the system under the CEMPC is more agile and steady when satisfying the customers' electric demand. On the other hand, the HRTO closely tracks the building temperature set-point while the proposed CEMPC achieves the zone tracking and basically keeps the building temperature within the acceptable range and close to its upper limit to increase system profitability. It is also observed that the CEMPC smoothly drives equipment to produce cooling and electricity, which contrasts sharply with the HRTO intensely regulating. For the microgrid, this frequent fluctuation of the electric power will threaten its safety. Besides, for both control schemes, the microturbine integrated with the absorption chiller always provides the electric and cooling baseload. The fuel cell and electric chiller are turned on to spread the load when the customers have great demands on electricity and cooling. The sketch of the CEMPC and HRTO input variables illustrated in Figure \ref{f7} confirms that the CEMPC gently regulates the manipulated inputs of the system. Moreover, Figure \ref{f8} shows the capacity states of the battery bank and chilled water storage unit under two control schemes. It is not difficult to see that the energy storage units can achieve the load shifting. The energy storage units are charged when the customers' demands are low and discharged when the customers require more electricity and cooling.

\subsubsection{Performance indices and analysis}
To further investigate the system performance under both control strategies, four performance indices are designed as follows:

(1) the summation of the absolute set-point tracking error in electric power $E_e$:
\begin{equation}
	\begin{aligned}
		E_e=&\sum_{i=1}^{N_{sim}} \|y_1(i) - y_{sp,1}(i)\|
	\end{aligned}
\end{equation}
where $N_{sim}$ represents the duration of simulation, $E_e$ represents the total error in tracking electric power and a smaller $E_e$ indicates better performance in meeting electric demand;

(2) the summation of the absolute zone tracking error in building temperature $E_t$:
\begin{equation}
	\begin{aligned}
		E_t=&\sum_{i=1}^{N_{sim}} \frac{1}{2}(\|y_2(i) - y_L(i)\| + \|y_2(i) - y_H(i)\| - \| y_H(i) - y_L(i)\|)
	\end{aligned}
\end{equation}
where $y_L$ and $y_H$ are the upper and lower limits on the acceptable range of building temperature and $y_L = y_{sp,2}-0.5$, $y_H = y_{sp,2}+0.5$, respectively; $E_t$ represents the accumulate error in tracking the temperature range and a smaller value of $E_t$ represents a superior performance in satisfying cooling requirements.

(3) the summation of the fuel consumption $E_f$:
\begin{equation}
	\begin{aligned}
		E_f=&\sum_{i=1}^{N_{sim}} (u_1(i) + u_2(i)) - k_{e,1}(k_{e,2}\Delta C_{soc} + k_{e,3}\Delta C_{sot})
	\end{aligned}
\end{equation}
where $E_f$ represents the system economic performance and consists of two parts: the first term is direct fuel consumption for generation, the second term is fuel consumption indirectly caused by the surplus of energy storage compared to their initial state; in the second term, $\Delta C_{soc}$ and $\Delta C_{sot}$ are the difference in capacity states of battery and cold tank; $k_{e,1} = 0.09807$, $k_{e,2} = 561.6$, $k_{e,2} = 106.1424$ are the conversion factor; of course, the smaller $E_f$, higher efficiency, is preferred.

(4) the overall performance of the IES $E_{oa}$:
\begin{equation}
	\begin{aligned}
		E_{oa}= E_e + E_t +E_f
	\end{aligned}
\end{equation}
where $E_{oa}$ reflects the system's overall performance and composed of the summation of $E_e$, $E_t$, $E_f$. Smaller $E_{oa}$ represents better performance.

\begin{table}[t] \small
	\centering
	\caption{Performance indices of the CEMPC and HRTO in Scenario 1, 2, 3}
	\label{t4}
	\renewcommand{\arraystretch}{1.2}
	\tabcolsep 10pt
	\begin{tabular}{ccccc} \hline
		Scenario 1 & $E_e$ & $E_t$ & $E_f$ & $E_{oa}$ \\ \hline
		CEMPC$_{(\delta=0.5)}$ & 691.80& 50.30& 665.60& 1407.7 (67.10\%) \\
		HRTO & 1413.40& 0& 684.70& 2098.1 (100\%) \\\hline
		Scenario 2 & $E_e$ & $E_t$ & $E_f$ & $E_{oa}$ \\ \hline
		CEMPC$_{(\delta=0.5)}$ & 1174.50& 327.10& 692.10& 2193.7 (83.10\%) \\
		HRTO & 1927.2& 0& 713.50& 2640.7 (100\%)\\\hline
		Scenario 3 & $E_e$ & $E_t$ & $E_f$ & $E_{oa}$ \\ \hline
		CEMPC$_{(\delta=0.5)}$ & 686.00& 78.70& 742.10& 1506.8 (81.80\%) \\
		HRTO & 1081.10& 0& 761.40& 1842.4 (100\%) \\ \hline
	\end{tabular}
\end{table}

The resulting performance indices of the CEMPC and HRTO in Scenario 1 are given in Table \ref{t4}. From Table \ref{t4}, we can see that the proposed CEMPC achieves a significant reduction in the electricity tracking error while improving the system economics compared to the HRTO. The overall performance $E_{oa}$ of the system under the CEMPC is thus improved by 32.9\% compared to under the HRTO in Scenario 1. On the other hand, we also note that the CEMPC drives the building temperature close to the upper limit of the acceptable range, which leads to a rise in the building temperature tracking error. This is the case because the building temperature is closer to the upper limit for saving more fuel, and the temperature tracking is relaxed to suppress the fluctuation of the electric power. It is worthwhile for the proposed CEMPC that gain more freedom to prioritize satisfying the electric requirements, considering the thermal comfort and the electric power having a significant impact on microgrids.

\begin{table}[t] \small
	\centering
	\caption{Computational time of the CEMPC and HRTO in Scenario 1, 2, 3}
	\label{t5}
	\renewcommand{\arraystretch}{1.2}
	\tabcolsep 4pt
		\begin{tabular}{cccccc} \hline
		CEMPC$_{(\delta=0.5)}$	& Long-term EMPC& Slow EMPC& Medium EMPC& Fast EMPC& Spent time per hour\\ \hline
			Scenario 1 & 154.1 s& 0.679 s& 0.824 s& 0.251 s& 997.07 s \\
			Scenario 2 & 163 s& 0.675 s& 0.817 s& 0.249 s& 999.1 s\\
			Scenario 3 & 158.4 s& 0.673 s& 0.83 s& 0.252 s& 1005.69 s\\ \hline
		\end{tabular}
	\begin{tabular}{ccccc}
	HRTO & Long-term optimization& Real-time optimization& Tracking MPC& Spent time per hour\\ \hline
	Scenario 1 & 132.7 s& 0.676 s& 3.93 s& 2982.58 s \\
	Scenario 2 & 144.2 s& 0.677 s& 3.95 s& 3008.51 s\\
	Scenario 3 & 142.9 s& 0.671 s& 3.9 s& 2971.03 s\\ \hline
\end{tabular}
\end{table}

Finally, to compare the computational complexity of the proposed CEMPC with the HRTO in Scenario 1, the average computational time per calculation and total time spent on computation per hour are presented in Table \ref{t5}. It is not difficult to see that the proposed CEMPC possesses a significantly shorter computation time of 66.6\% compared with the HRTO, demonstrating that the CEMPC is a practical scheme for coordinating IESs and the multi-time-scale decomposition alleviates the computational burden.

\subsection{Results of IES under varying external conditions}

\begin{figure}[t]
	\centering
	\includegraphics[width=0.62\hsize]{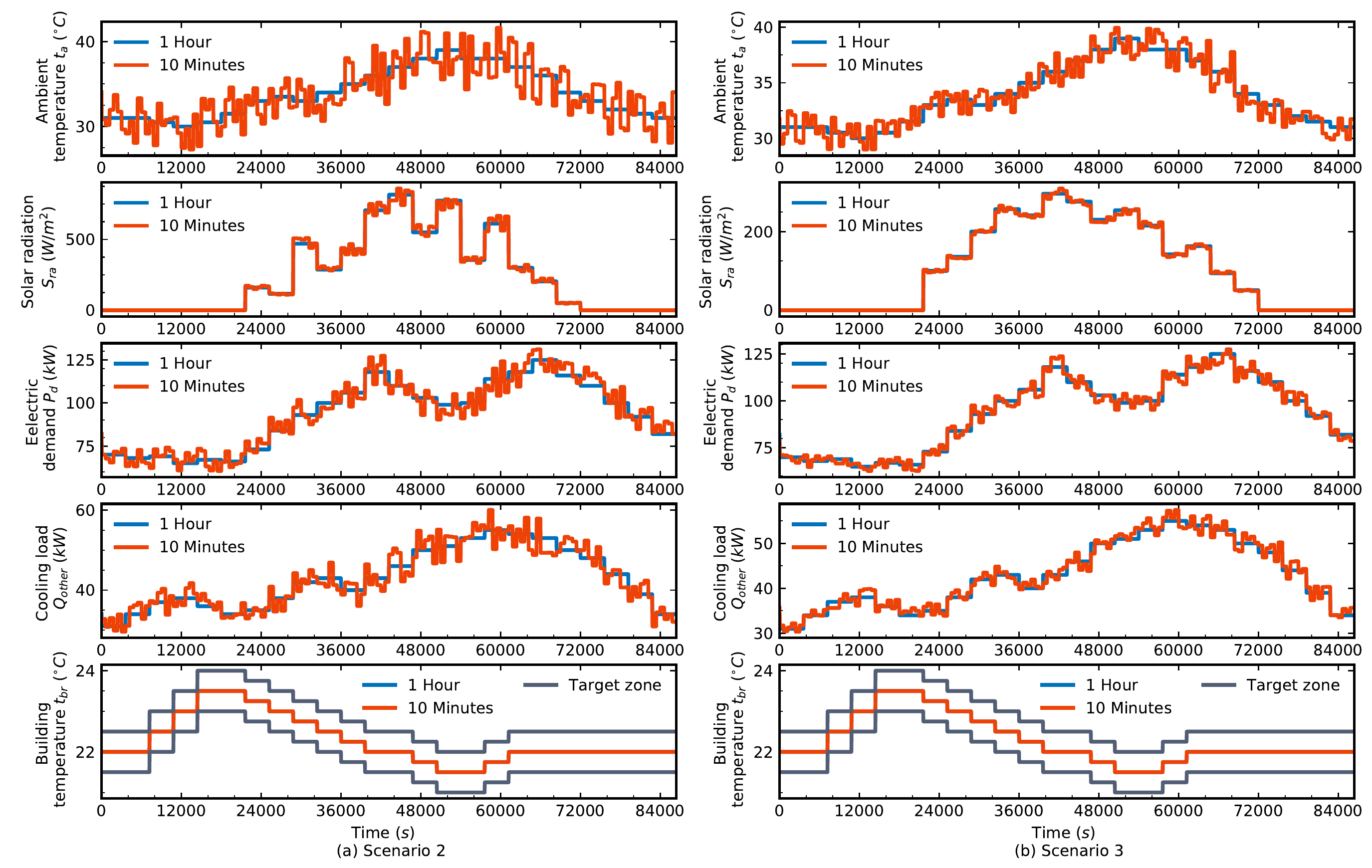}
	\caption{Varying external conditions (a) Drastic variations in external conditions (Scenario 2), (b) Insufficient solar energy conditions (Scenario 3)}
	\label{f9}
\end{figure}

Given the uncertainty of external conditions, this subsection discusses the performance of the control schemes under drastic variations in external conditions (Scenario 2) and insufficient solar energy (Scenario 3), as shown in Figure \ref{f9}. In Scenario 2, the fluctuating ambient temperature and solar radiation occur in the long-term forecast. At the same time, the actual environmental conditions and customers' demands fluctuate randomly within a range of $\pm 10\%$ around the long-term forecast. In Scenario 3, the peak of solar radiation is 300$W/m^2$ and the actual environment and demand curve vary randomly within $\pm 5\%$ of the long-term curve. In the proposed CEMPC, the relaxation value $\delta$ is also chosen as 0.5.

\subsubsection{Resulting performance}
\begin{figure}[t]
	\centering
	\includegraphics[width=0.95\hsize]{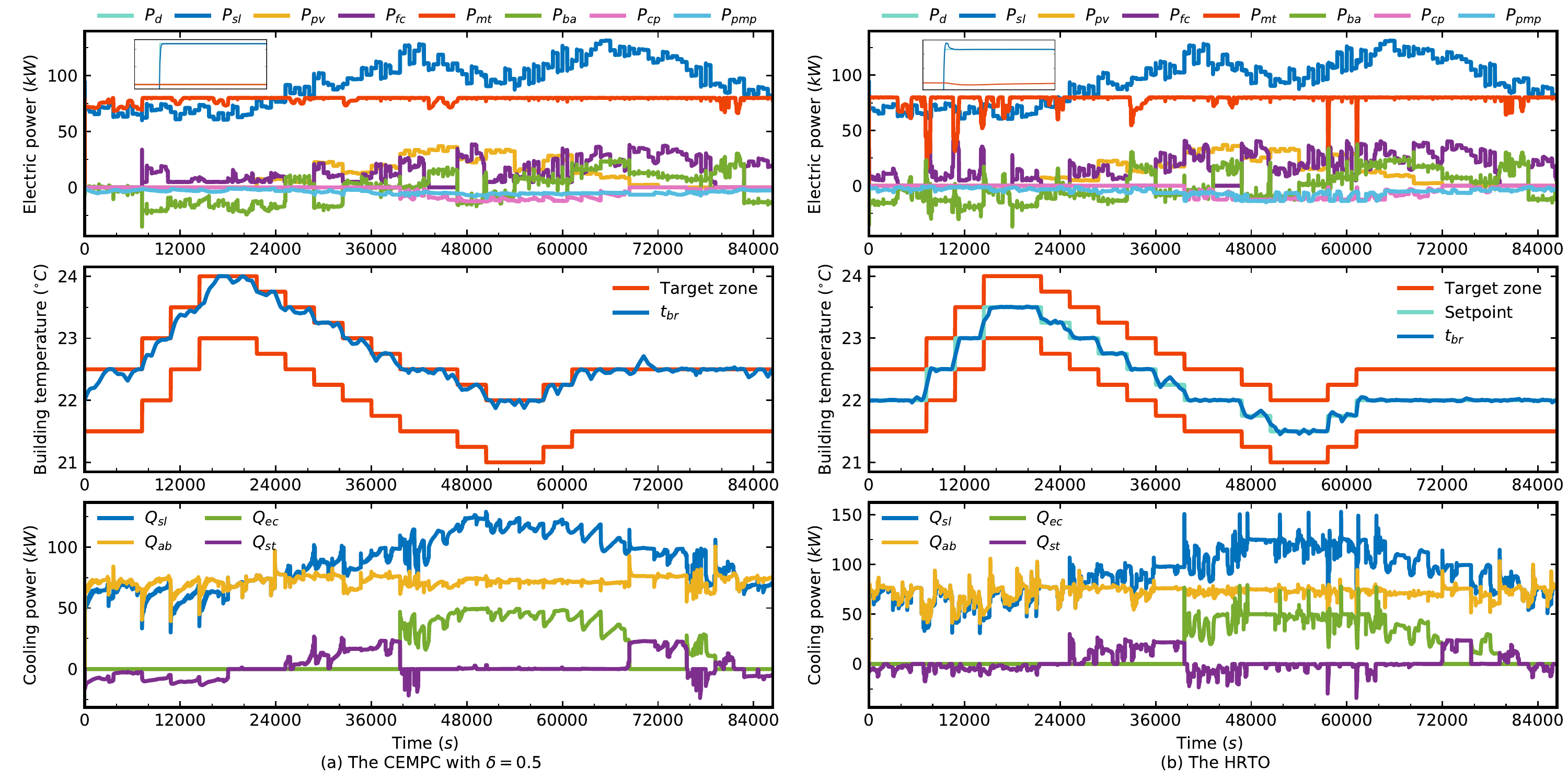}
	\caption{The outputs of the IES and power generation in each operating unit under (a) CEMPC, (b) HRTO in Scenario 2}
	\label{f11}
\end{figure}
\begin{figure}[h]
	\centering
	\includegraphics[width=0.95\hsize]{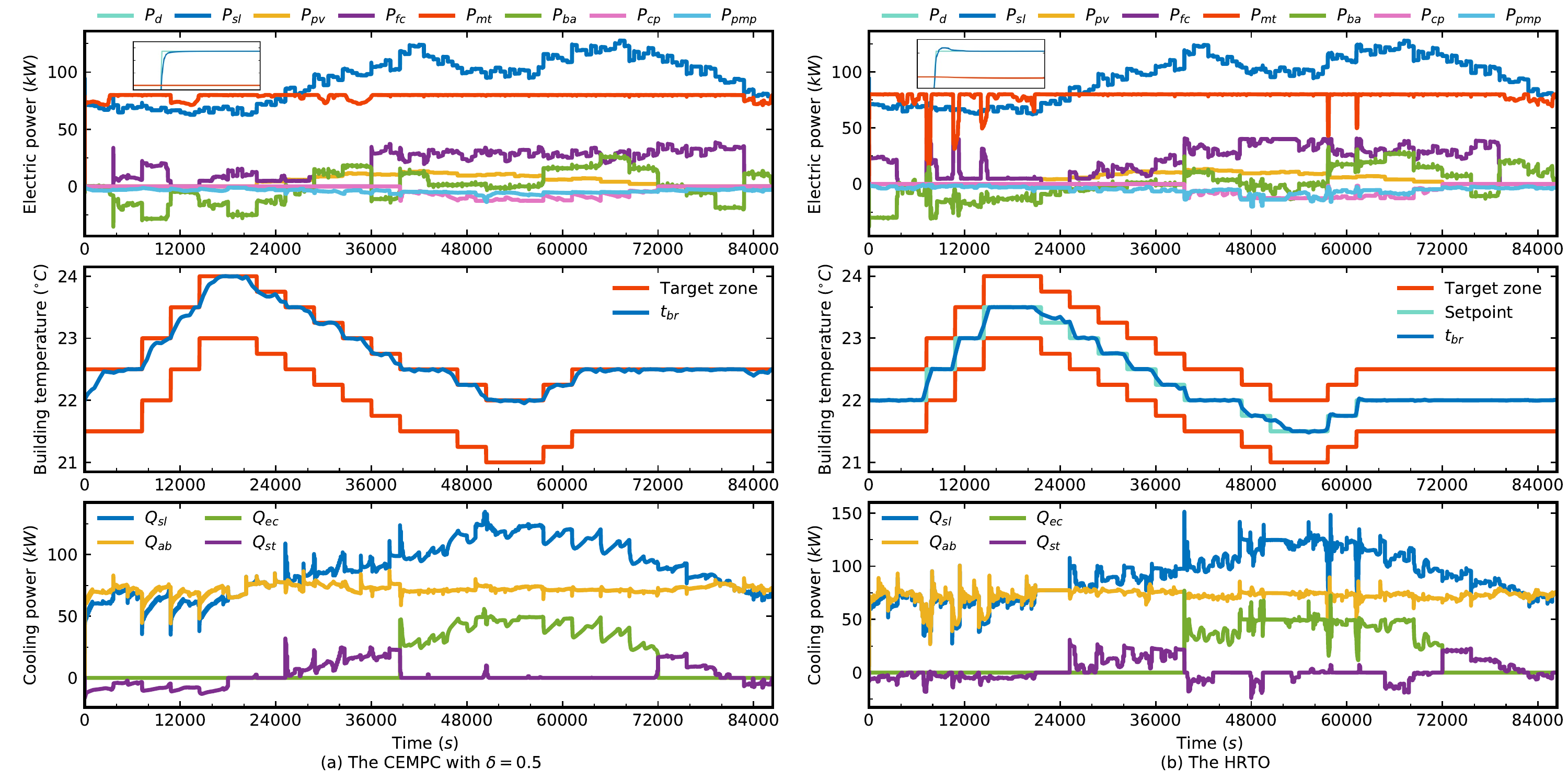}
	\caption{The outputs of the IES and power generation in each operating unit (a) CEMPC, (b) HRTO in Scenario 3}
	\label{f15}
\end{figure}

The resulting performances of the control schemes in Scenario 2 are shown in Figure \ref{f11}. As can be seen from the figure, the outputs curves under both the CEMPC and HRTO become steep compared with the curves in Scenario 1, and the system power output and building temperature also fluctuate obviously. However, the proposed CEMPC can still improve the tracking performance of electric power and alleviate the oscillations of the output power in each operating unit. This achievement results in a slight exacerbation of the building temperature variation within an acceptable range.

Figure \ref{f15} displays the system behavior of the two controllers in Scenario 3. Comparing Scenario 1, 2, the proposed CEMPC and HRTO drive the system smoothly due to reduced external disturbances. The CEMPC is still superior to the HRTO from the perspective of meeting electric requirements. Besides, the operating units are controlled by the CEMPC in a temperate way while maintaining a smooth transition of building temperature within the acceptable range.

\subsubsection{Analysis of the performance indices}
Table \ref{t4} also exhibits the performance indices of the two control schemes in Scenarios 2 and 3, respectively. It can be found that although the tracking performance of the building temperature under the CEMPC is degraded in Scenario 2, the CEMPC can still significantly reduce the electricity tracking error and improve the system's economic performance compared with the HRTO. Consequently, the CEMPC improves the overall performance $E_{oa}$ of the system by 16.9\% under drastic external conditions. In Scenario 3, the system's fuel consumption rises substantially due to insufficient solar energy. Also, since the external disturbances become weaker, the load tracking performance under the two controllers has been slightly improved compared with those in Scenarios 1 and 2. The proposed CEMPC remains evidently superior to the HRTO in terms of the economics, electric load tracking and overall performance.

The computational time of the CEMPC and HRTO under Scenarios 2 and 3 are given in Table \ref{t5} as well, which shows the computational time of the CEMPC and HRTO almost stays the same as in Scenario 1 and the CEMPC is still much quicker than the HRTO in solving the optimization problem.

\subsection{Investigation of changing zone relaxation value $\delta$}
It has been detected from the previous simulations that the CEMPC will drive the building temperature towards the upper limit of the acceptable range, which is also the limit of $\delta$ when setting the relaxation variable $\delta$ equals the acceptable range. This situation will sometimes cause the building temperature to slightly exceed the acceptable range's upper limit. Therefore, it is necessary to investigate how the relaxation value $\delta$ impacts on the control performance of the CEMPC and how the CEMPC performs if the customers request a strict set-point tracking for the building temperature. This subsection presents simulation results when $\delta$ takes different values and the set-point tracking for the building temperature is required.

\subsubsection{Zone tracking for building temperature with different $\delta$}
\begin{figure}[t]
	\centering
	\includegraphics[width=0.98\hsize]{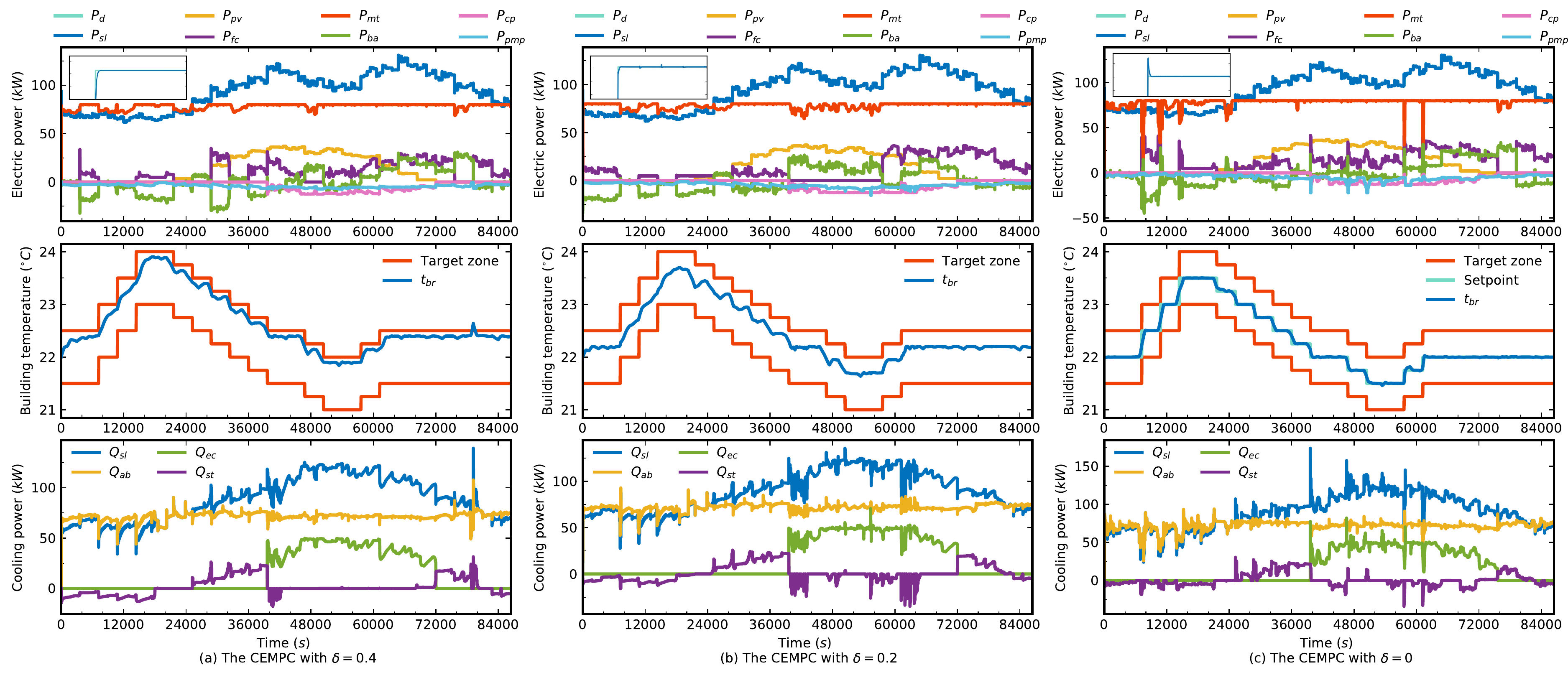}
	\caption{The outputs of the IES and power generation in each operating unit under CEMPC with (a) $\delta=0.4$, (b) $\delta=0.2$, (c) $\delta=0$ in Scenario 1}
	\label{f19}
\end{figure}
The simulation results of the CEMPC with $\delta = 0,0.2,0.4$ in Scenario 1 are shown in Figure \ref{f19}. It is not difficult to see that with the decrease of $\delta$, the tracking performance of the on electric power gradually deteriorates, and the building temperature progressively comes to the set value. At the same time, the power outputs of each operating unit become sharper, which indicates that the operating units are frequently and intensely regulated. By setting $\delta = 0$ the CEMPC achieves the building temperature set-point tracking, in this case the behavior of the system is similar to the HRTO.

\begin{table}[t] \small
	\centering
	\caption{Performance indices of the CEMPC with different $\delta$ in Scenario 1, 2, 3}
	\label{t6}
	\renewcommand{\arraystretch}{1.2}
	\tabcolsep 10pt
	\begin{tabular}{ccccc} \hline
		Scenario 1& $E_e$ & $E_t$ & $E_f$ & $E_{oa}$ \\ \hline
		CEMPC$_{(\delta=0.4)}$ & 807.3& 39.9& 665.6& 1512.8 \\
		CEMPC$_{(\delta=0.3)}$ & 838.6& 0& 670.6& 1509.2 \\
		CEMPC$_{(\delta=0.2)}$ & 825.5& 0& 673.1& 1498.6 \\
		CEMPC$_{(\delta=0.1)}$ & 1028.7& 0& 676.2& 1704.9 \\ \hline
		Scenario 2& $E_e$ & $E_t$ & $E_f$ & $E_{oa}$ \\ \hline
		CEMPC$_{(\delta=0.4)}$ & 1121.4& 11.6& 695.7& 1828.7 \\
		CEMPC$_{(\delta=0.3)}$ & 1192.9& 0& 699.8& 1892.7 \\
		CEMPC$_{(\delta=0.2)}$ & 1218.5& 0& 698.3& 1916.8 \\
		CEMPC$_{(\delta=0.1)}$ & 1239.7& 0& 703.4& 1943.1 \\ \hline
		Scenario 3& $E_e$ & $E_t$ & $E_f$ & $E_{oa}$ \\ \hline
		CEMPC$_{(\delta=0.4)}$ & 690.2& 0.9& 743.8& 1434.8 \\
		CEMPC$_{(\delta=0.3)}$ & 834.5& 0& 750.9& 1585.4 \\
		CEMPC$_{(\delta=0.2)}$ & 744.5& 0& 751.1& 1495.6 \\
		CEMPC$_{(\delta=0.1)}$ & 901.7& 0& 756.3& 1658 \\ \hline
	\end{tabular}
\end{table}
\begin{figure}[t]
	\centering
	\includegraphics[width=0.95\hsize]{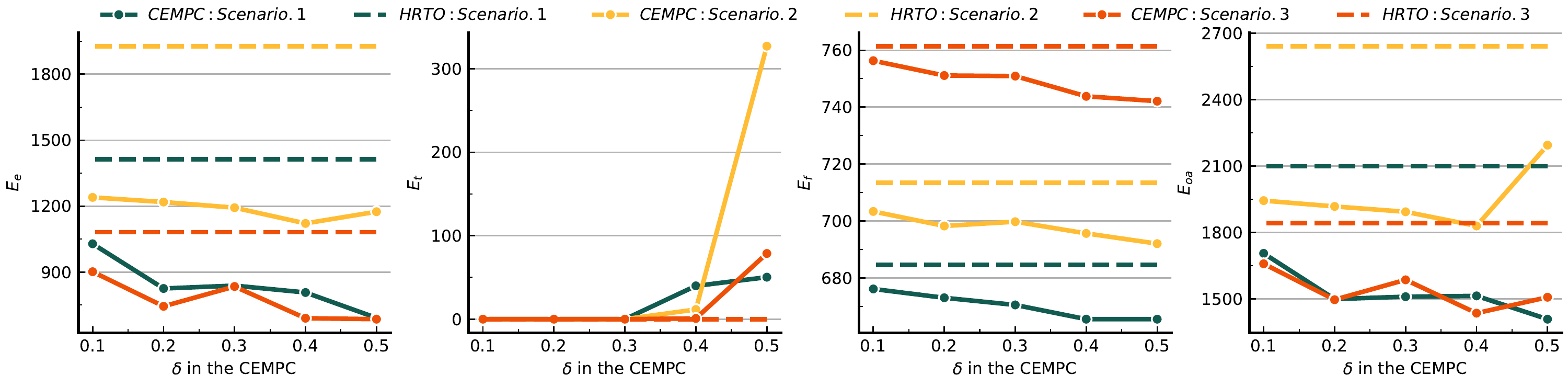}
	\caption{The performance indices of HRTO and CEMPC with different $\delta$}
	\label{f24}
\end{figure}
The evolution of performance indices under the CEMPC with $\delta = 0.1,0.2,0.3,0.4$ in different scenarios is displayed in Table \ref{t6} and sketched in Figure \ref{f24}. It can be seen that the electricity tracking error $E_e$, the economic performance $E_f$ and the overall performance $E_{oa}$ under the CEMPC much better than under the HRTO given in Table \ref{t4} no matter how large $\delta$ is chosen, and $E_e$, $E_f$ and $E_{oa}$ generally decline with $\delta$ increase. Compared to the HRTO, the CEMPC can effectively reduce $E_e$, $E_f$ and $E_{oa}$ and keep $E_t=0$ even if $\delta=0.1$, which demonstrates that the electricity tracking performance, the system economic performance and the overall performance can be improved by the zone tracking for the building temperature. Meanwhile, $E_t$ remains around 0 until $\delta=0.3$, then $E_t$ increases with $\delta$.

\subsubsection{Set-point tracking for building temperature}
In the situation that the building temperature is required to track precisely, $\delta$ is set as 0 in the CEMPC, and the CEMPC becomes the set-point tracking form. In this case, the simulation results of the CEMPC in Scenario 1 have been shown in Figure \ref{f19} and discussed. The performance indices of the CEMPC and HRTO in different scenarios are displayed in Table \ref{t7}. It should be noted that in this case the parameters $y_L$ and $y_H$ in the performance index $E_t$ are set as $y_L=y_H=y_{sp,2}$ to reflect the absolute set-point tracking error in the building temperature. From Table \ref{t7}, we can see that all of the indices of the CEMPC are better than the HRTO in Scenario 1, 2, in which the overall performances $E_{oa}$ of the system are improved by 5\% via the CEMPC. In Scenario 3, the CEMPC achieves much better $E_t$ and $E_f$ than the HRTO at the expense of slight deterioration in $E_e$. Consequently, the overall performance $E_{oa}$ of the CEMPC remains superior to the HRTO. 

\begin{table}[t] \small
	\centering
	\caption{Performance indices of the CEMPC and HRTO in case of set-point tracking in Scenario 1, 2, 3}
	\label{t7}
	\renewcommand{\arraystretch}{1.2}
	\tabcolsep 10pt
	\begin{tabular}{ccccc} \hline
		Scenario 1 & $E_e$ & $E_t$ & $E_f$ & $E_{oa}$ \\ \hline
		CEMPC$_{(\delta=0)}$& 1344& 1273.4& 681.5& 3299 (95.00\%) \\
		HRTO& 1413.40& 1374& 684.70& 3472.1 (100\%) \\\hline
		Scenario 2 & $E_e$ & $E_t$ & $E_f$ & $E_{oa}$ \\ \hline
		CEMPC$_{(\delta=0)}$& 1822.00& 1683.5& 710.80& 4216.3 (95.00\%) \\
		HRTO& 1927.2& 1796.5& 713.50& 4437.2 (100\%) \\\hline
		Scenario 3 & $E_e$ & $E_t$ & $E_f$ & $E_{oa}$ \\ \hline
		CEMPC$_{(\delta=0)}$& 1177.00& 1594.5& 759.50& 3531.1 (96.00\%) \\
		HRTO& 1081.10& 1836.2& 761.40& 3678.7 (100\%) \\ \hline
	\end{tabular}
\end{table}
\begin{table}[t] \small
	\centering
	\caption{Computational time of the CEMPC in case of set-point tracking in Scenario 1, 2, 3}
	\label{t8}
	\renewcommand{\arraystretch}{1.2}
	\tabcolsep 4pt
	\begin{tabular}{cccccc} \hline
		CEMPC$_{(\delta=0)}$& Long-term EMPC& Slow EMPC& Medium EMPC& Fast EMPC& Spent time per hour\\ \hline
		Scenario 1 & 135.9 s& 0.675 s& 0.828 s& 0.254 s& 985.95 s \\
		Scenario 2 & 146.5 s& 0.681 s& 0.822 s& 0.252 s& 990.43 s \\
		Scenario 3 & 145.9 s& 0.679 s& 0.83 s& 0.257 s& 1002.37 s \\ \hline
	\end{tabular}
\end{table}
With respect to the computational complexity, the computational time of the CEMPC with set-point tracking in different scenarios is given in Table \ref{t8}, which shows that the computational complexity of the CEMPC is basically consistent with the data given in Table \ref{t5} and the CEMPC is more efficient than the HRTO. These results further demonstrate that the multi-time-scale decomposition in the CEMPC not only boosts the computational efficiency, but is also capable of addressing the time-scale multiplicity and couplings in IESs to improve the system performance.

\section{Conclusions}
In this work, we proposed a novel composite economic MPC with zone tracking for a stand-alone integrated energy system to address the needs of customers' load and system economics based on multi-time-scale decomposition. Extensive simulations were carried out to compare the performance of the proposed CEMPC and the HRTO. The simulations show that the proposed CEMPC provides superior electricity tracking, system profitability and overall performance while maintaining the building temperature within an acceptable range or at the desired set-point. On the other hand, it is demonstrated that the CEMPC based on multi-time-scale decomposition is able to match the time-scale multiplicity featured in the system dynamics and improve the computational efficiency, making the CEMPC superior to the HRTO. Further, the CEMPC allows for achieving different control objectives when tuning the relaxation variable in zone tracking, which offers the flexibility and reliability to regulate the operating performance according to actual environmental conditions and customers' requirements.

\section{Acknowledgment}
This work was supported by National Nature Science Foundation of China (51936003); National Key R\&D Program of China (2018YFB1502901); China Scholarship Council. The second author X. Yin would like to acknowledge the financial support from Nanyang Technological University (Start-Up Grant) and Ministry of Education Singapore (MOE AcRF Tier 1 Seed Funding).

\end{document}